%% file: spyglass4.tex
\documentclass[twocolumn]{aastex63}

\usepackage{listings}
\usepackage{mathtools}
\usepackage{longtable}
\usepackage{wasysym}
\usepackage{booktabs}
\usepackage[caption=false]{subfig}
\shorttitle{Recent Star Formation Within 1 kpc}
\shortauthors{Kerr et al.}

\begin{document}

\title{SPYGLASS. IV. New Stellar Survey of Recent Star Formation within 1 kpc}

\correspondingauthor{Ronan Kerr}
\email{rmpkerr@utexas.edu}

\author[0000-0002-6549-9792]{Ronan Kerr}
\affiliation{Department of Astronomy, University of Texas at Austin\\
2515 Speedway, Stop C1400\\
Austin, Texas, USA 78712-1205\\}

\author[0000-0001-9811-568X]{Adam L. Kraus}
\affiliation{Department of Astronomy, University of Texas at Austin\\
2515 Speedway, Stop C1400\\
Austin, Texas, USA 78712-1205\\}

\author{Aaron C. Rizzuto}
\affiliation{Department of Astronomy, University of Texas at Austin\\
2515 Speedway, Stop C1400\\
Austin, Texas, USA 78712-1205\\}



\begin{abstract}

Young stellar populations provide a powerful record that traces millions of years of star formation history in the solar neighborhood. Using a revised form of the SPYGLASS young star identification methodology, we produce an expanded census of nearby young stars (Age $<50$ Myr). We then use the HDBSCAN clustering algorithm to produce a new SPYGLASS Catalog of Young Associations (SCYA), which reveals 116 young associations within 1 kpc. More than 25\% of these groups are largely new discoveries, as 20 are substantively different from any previous definition, and 10 have no equivalent in the literature. The new associations reveal a yet undiscovered demographic of small associations with little connection to larger structures. Some of the groups we identify are especially unique for their high transverse velocities, which can differ from the solar velocity by 30-50 km s$^{-1}$, and for their positions, which can reach up to 300 pc above the galactic plane. These features may suggest a unique origin, matching existing evidence of infalling gas parcels interacting with the disk ISM. Our clustering also suggests links between often-separated populations, hinting to direct structural connections between Orion Complex and Perseus OB2, and between the subregions of Vela. The $\sim$30 Myr old Cepheus-Hercules association is another emerging large-scale structure, with a size and population comparable to Sco-Cen. Cep-Her and other similarly-aged structures are also found clustered along extended structures perpendicular to known spiral arm structure, suggesting that arm-aligned star formation patterns have only recently become dominant in the solar neighborhood.

\end{abstract}

\


\section{Introduction} \label{sec:intro}

Most young stars are located in clusters and associations, stellar populations left behind after the dispersal of their natal cloud \citep[][]{lada03,Krumholz19}. By preserving elements of the dynamics and history of their star-forming environments, these populations provide a powerful record of recent star formation which can be detected for tens of millions of years after their formation \citep{Mamajek16, Krause20}. Detailed studies of young populations therefore have the unique ability to reconstruct entire star formation histories for recent events, using dynamics to trace the locations of stars back in time, and ages to determine the location of stars and their subsequent environments at the time of formation \citep[e.g.,][]{Galli21,MiretRoig22, Kerr22b, Kerr22a}. The results can reveal star formation patterns at a range of scales, from molecular clouds to spiral arms \citep[e.g.,][]{Pecaut16, Pang21, PantaleoniGonzalez21, Zucker22}. 

On the scale of individual associations, young populations can be used to trace star formation patterns that either last too long to understand in full from observations of active star-forming regions, or are too rare or brief for comparable active sites to be readily available. Association-level studies have recently been used to provide evidence for processes such as triggered star formation, where star formation is initiated by an external force such as a supernova, or sequential star formation, where star formation propagates across a molecular cloud, with each star-forming event producing the feedback which initiates the formation of the next generation \citep[e.g.,][]{Elmegreen77, Kerr21, Nony21, Pang21}. On larger scales, young populations can be used to trace galactic spiral arm structure \citep{Zucker22}. The recent discoveries of the Radcliffe Wave and the Split reveal structures largely aligned with the pitch of our spiral arms, as do structures marked by O and B stars \citep{Lallement19, Alves20, PantaleoniGonzalez21}. However, much less is known about the distribution of older, gas-poor structures with ages $\gtrsim20$ Myr outside of the nearest 100 pc, especially less populated structures that lack O and B stars \citep{deZeeuw99, Mamajek16, Zucker22}. Broad scale is therefore an asset for studies of young populations, enabling the emergence of large-scale patterns from disparate nearby associations. With the emergence of increasingly comprehensive simulations of star formation, studies of young associations will become increasingly important for testing star formation models \citep[e.g.,][]{Grudic20, Guszejnov22}.  

Until recently, our view of nearby stellar populations has been limited to only the nearest and most substantial populations \citep[][]{deZeeuw99, Mamajek16}. Within 50-100 pc, the detection of associations has generally relied on stars with strong youth indicators such as TW Hydrae and $\beta$ Pic, around which associations can be built by searching for nearby stars with common motion \citep{Kastner97, Barrado99, Zuckerman01, Gagne18}. For more distant populations, the short-lived and therefore necessarily young O and B stars assume the role of signposts, and can be grouped together into comoving OB Associations. While this approach can reliably detect large associations such as Perseus OB3, the Orion Nebula Complex, and Sco-Cen \citep[e.g.,][]{deZeeuw99}, small associations often lack O and B stars, severely limiting the range of structures over which this approach is useful. This has resulted in more distant low-mass structures being largely unexplored until very recently \citep[e.g., see][]{Kounkel19, Kerr21, Prisinzano22}. 

The Gaia survey has revolutionized the detection of young stellar populations by providing nearly 2 billion stars with accurate space-velocity coordinates and photometric measurements \citep[][]{GaiaDR218, GaiaEDR321}. This not only allows for the detection of co-moving structures, but also the suppression of the field background by detecting and isolating stars with photometry consistent with youth. Recent surveys have already identified thousands of stellar populations and clusters across a wide range of ages \citep[e.g.,][]{Sim19,Kounkel19,CantatGaudin20,Hunt23}.
While the populations discovered by these studies were numerous, their broad focus in age reduced the visibility of young structures. \citet{Zari19} used a photometrically-limited sample to identify patterns in the distribution of young stellar populations. This work however did not cluster the distribution of young stars into groups, meaning that the extents of potential populations were left undefined. Surveys of individual populations occasionally include age-limited samples of young stars and the clustering of substructures \citep[e.g.,][]{Zari18, CantatGaudin19}, however, until very recently, there were no all-sky surveys specifically targeting young stellar populations in Gaia. 

The SPYGLASS program (Stars with Photometrically Young Gaia Luminosities Around the Solar System), which this work is a part of, was designed to complement the existing research on young stellar populations by creating a spatially unbiased survey of young stellar populations which both robustly assesses the youth of potential members and provides well-defined extents and membership lists for the populations that emerge from that sample. The first paper in this series, \citet{Kerr21} (hereafter SPYGLASS-I), outlined our Bayesian framework for the detection of young stars and performed a relatively conservative clustering analysis on a Gaia DR2-based sample. The search focused on populations under 50 Myr old within 333 pc of the sun, revealing 27 top-level associations with numerous subclusters, many of which were either little-known or completely absent from the literature. While SPYGLASS-I considerably expanded our record of nearby associations, it was relatively conservative in its clustering analysis, using quality cuts to avoid the inclusion of populations in the background of dense molecular clouds where reddening corrections occasionally allowed the erroneous detection of subgiants as young because reddening was underestimated. 

The recent release of Gaia Data Release 3 (DR3) provides a new opportunity to deepen our survey of these nearby populations. The sample represents a significant improvement over DR2, improving the precision of parallaxes by 30\%, proper motions by a factor of 2, and greatly reducing the systematic errors for both of those measures \citep{GaiaEDR321}. The EDR3 sample, which is largely identical to the DR3 sample with the exception of radial velocities \citep{GaiaDR322}, has already been used by \citet{Prisinzano22} for an expansive survey of young stellar populations, revealing 354 associations under 10 Myr old within a radius of 1.5 kpc. This result demonstrates the power of this updated Gaia sample to reveal stellar structures far beyond the 333 pc radius in SPYGLASS-I. However, that work's focus on populations younger than 10 Myr and lack of spatially-dependent reddening and extinction corrections motivates new work that covers a wider range of ages and includes reddening as a core component of stellar youth assessment.

Through a Gaia DR3 update to our SPYGLASS young star and association detection framework, we can take advantage of the quality improvements of DR3 while better optimizing our young star detection algorithm, and adapting vetting methods to exclude false groups while including tenuous stellar populations. In this paper, we outline our expanded survey of young stellar populations in the solar neighborhood, both improving our sensitivity to young stellar populations relative to SPYGLASS-I and widening the survey to 1 kpc. In Section \ref{sec:data} we outline the new Gaia DR3 dataset that we make use of. We describe updates to our SPYGLASS methodology in Section \ref{sec:methods}, which refines our identification of young stars, and provide revised clustering results and cluster vetting techniques in Section \ref{sec:populations}, which improve our sensitivity to young associations. We then outline a new technique for computing cluster membership probabilities in Section \ref{sec:pmem}, before providing basic information on the groups we detect in Section \ref{sec:results}. We then provide an overview of some broad features of the groups we detect in Section \ref{sec:discussion}, before concluding in Section \ref{sec:concl}. 

\section{Data} \label{sec:data}

The recent publication of Gaia Data Release 3 (DR3) has significantly improved the astrometric and photometric quality of measurements from the Gaia spacecraft compared to DR2, which was used in SPYGLASS-I \citep{GaiaMission,GaiaDR218,GaiaDR322}. We therefore use that updated dataset for this paper. Our initial data download from DR3 was much less restrictive compared to the SPYGLASS-I sample to reflect our desire to produce maximally complete populations for any associations we find. We required only that each star has a 5-parameter Gaia astrometric solution and a valid $G$ magnitude, and that the star is within 1 kpc according to the \citet{BailerJones21} geometric distances (see Section \ref{sec:distances}). The resulting sample contains approximately 94 million stars. This dataset  provides a maximally complete sample capable of producing deep coverage of nearby populations. 

While this unrestricted sample is necessary for complete demographic studies, not all objects included are well-suited for youth assessment and the group detection that it enables. We therefore produce a separate restricted sample of stars with quality diagnostics that are well-suited for youth assessment. To enable the quality restriction of the sample, we included a series of quality parameters in our data download. This allowed us to generate a series of quality flags, which are based on the following inequalities outlined in SPYGLASS-I:
\begin{equation}\label{eqn:astroflag}
\begin{multlined}
u < 1.2 \times \max[1, \exp(-0.2(G - 19.5))]
\end{multlined}
\end{equation}

\begin{equation}\label{eqn:photflag}
\begin{multlined}
1.0 + 0.015(G_{BP} - G_{RP})^2 < E < \\1.3 + 0.037(G_{BP} - G_{RP})^2
\end{multlined}
\end{equation}
\begin{equation}\label{eqn:parflag}
\pi/\sigma_\pi>5
\end{equation}

\noindent where $u$ is the unit weight error, defined as $u = \sqrt{\chi^2/\nu}$, with $\chi^2$ being the goodness of fit of the single-star astrometric solution\footnote{{\tt astrometric\_chi2\_all} in the Gaia archive}, and $\nu$ being the number of observations used in that solution\footnote{{\tt astrometric\_n\_good\_obs\_al} in the Gaia archive}. $G$, $G_{RP}$, and $G_{BP}$ refer to the Gaia magnitudes, and $E$ is the BP/RP Flux Excess Factor\footnote{{\tt phot\_bp\_rp\_excess\_factor} in the Gaia archive}, which is an indicator of flux anomalies between the gaia $G$ band and the $G_{RP}$ and $G_{BP}$ bands. The inequalities \ref{eqn:astroflag} and \ref{eqn:photflag} can be converted to astrometric and photometric goodness flags, respectively, where stars that pass the inequalities have values set to 1, and stars that fail them have values of 0. Stars that fail these cuts are typically in crowded fields or occasionally close binaries where it is more difficult to disentangle the astrometry and photometry of separate sources. 

Alternative cuts using the Renormalized Unit Weight Error (RUWE) have recently become popular for vetting astrometric solutions in place of the cut for $u$, particularly the requirement that RUWE $<$ 1.4 \citep{RUWELindegren18}. This cut is most often used specifically to identify probable binaries \citep{Bryson20}, and while binaries do provide challenges to youth assessment, they are also a core component of all young associations, and are included in our model of the solar neighborhood introduced in Section \ref{sec:methods}. The requirement that RUWE $<$ 1.4 is also more than twice as restrictive as the cut on $u$, removing over 3.3 million stars out of the 54 million that exist before the astrometric cut. Furthermore, recent work by \citet{Fitton22} has shown that RUWE is also increased by the presence of protoplanetary disks, which are a common feature of young associations. We therefore conclude that the possible quality improvements that this RUWE cut provides do not justify the detrimental effects on the completeness of young populations, particularly at the youth probability calculation stage. We therefore do not apply this restriction for the initial selection of stars. Applying this cut is however useful for improving our detection of young associations in velocity space, a choice that we discuss in Section \ref{sec:clustering}. 

Inequality \ref{eqn:parflag} provides an additional quality check on parallax specifically, with a more permissive limit of 5 compared to the value of 10 chosen in SPYGLASS-I. This looser restriction is common in papers using Gaia data \citep[e.g.][]{Arenou18,GaiaRVsKatz19}, and its use reflects improvements made in Section \ref{sec:starstats}, which greatly improve our handling of parallax uncertainties. While additional restrictions related to distance uncertainty will be necessary for identifying young groups, further parallax restrictions are no longer required for accurate assessments of youth. Finally, we dropped the requirement that ${\tt visibility\_periods\_used}>8$ used in SPYGLASS-I, as no stars fail that restriction in Gaia DR3. Stars that passed these quality restrictions were admitted to the main quality-restricted dataset we used for analysis, which contains nearly 53 million stars within 1 kpc. 


\subsection{Distances} \label{sec:distances}

While inverting Gaia parallaxes provides an accurate distance measurement in the near field, when expanding our search to a distance of 1 kpc the uncertainties in these measurements become much more significant and asymmetrical, reducing the accuracy of this distance calculation method. To improve these results, \citet{BailerJones21} uses a series of priors to refine the distance measurements relative to results from raw Gaia parallaxes. This work provides both geometric distances, which use a direction-dependent prior in addition to Gaia parallaxes, and photogeometric distances, which include an additional color-magnitude prior which favors distances which produce positions in color-magnitude space consistent with expectations. Both of these measurement methods have been shown to produce a robust improvement over inverted parallaxes, especially for sources with larger fractional uncertainties \citep{LutzKelker73, BailerJones21}. We experimented with both distance calculation methods, running our full young star identification pipeline and HDBSCAN clustering routine on both datasets. We found limited visually identifiable differences between the two, with locally larger radial scatter in the photogeometric distances for some associations. While these differences were very subtle, we selected the geometric distances for the purposes of this project, despite the higher accuracy of the photogeometric distances reported in \citet{BailerJones21}. Since our detection of young stars is based on a Bayesian framework that uses distances and magnitudes, the use of distances with their own priors on magnitude may introduce undesired artifacts. This suspicion appears to be reflected in the subtly wider scatter in some associations during our testing, which may be the result of reddening anomalies manipulating and distorting the priors for the \citet{BailerJones21} photogeometric distances. Since both reddening and photometric youth are likely to skew generalized photometric priors, it would not be surprising to see less accurate photogeometric distances in these environments despite their higher accuracy relative to geometric distances across the rest of the sky. 

\section{Young Star Survey} \label{sec:methods}

Our methods for identifying young stars and associations, as well as computing their basic properties, closely follow the methods of SPYGLASS-I. However, we do provide some minor updates that improve performance. These changes include updates which both reflect the new Gaia DR3 photometric system and refine our young star identification methods. 

\subsection{Model Generation}

Like in SPYGLASS-I, our methods necessitate a model of stellar populations to compare with the Gaia sample we gathered in Section \ref{sec:data}. This requires model stars that are representative of the solar neighborhood, and capture the diversity in age, mass, metallicity, and binarity that exists. All of these factors modify luminosity and thereby affect the probability of youth for possible young stars. Most components of our model generation directly follow SPYGLASS-I, which can be referenced for further detail on our methods. 

We assumed constant star formation over the age of the solar neighborhood from 1 Myr to 11.2 Gyr \citep{Binney00}. We then sampled this distribution uniformly in log-space to ensure strong model coverage on the pre-main sequence, and accounted for the subsequent overselection of young stars in the model using a prior in our youth probability calculation (see Equation \ref{eqn:pyoung}). Metallicities are drawn from the probability distribution provided by the GALAH survey \citep{GALAHDR2Buder18, Hayden19}, in the smoothed form shown in SPYGLASS-I. While metallicity does have some dependence on the galactic Z coordinate over a 1 kpc scale, associations are rare beyond about 100 pc from the galactic plane \citep[e.g.,][]{Bobylev16}, so relatively few associations are likely to be affected by significant metallicity variations. The masses of single or primary system components are then drawn from the \citet{Chabrier05} system IMF, with possible companions added according to the binary and triple system rate curves from SPYGLASS-I, which are based on the multiplicity rates and higher-order multiplicity behaviors from \citet{Duchene13}. We also used the same mass ratio distribution for companions that was previously used in SPYGLASS-I, which were based on the power law distributions from \citet{Kraus11} and \citet{Rizzuto13}. System separations were also generated to assess whether the binaries are resolved. Unresolved binaries are visible as a single source, requiring that the photometry of the components be merged in the model, while resolved systems remain separate. The separations we use follow the distribution from \citet{Raghavan10}, which was also used in SPYGLASS-I.

We then generated model photometry by interpolating photometric observables from isochrones according to the randomly-generated stellar properties described above. Like in SPYGLASS-I, we used PARSEC isochrones \citep{PARSECChen15} to generate this model photometry, adopting the revised DR3 photometric system \citep{Riello21} and updating our isochrone grids to reflect that. Following SPYGLASS-I, we based our grid density on the rate of stellar evolution, requiring that each slice in age and mass must contain at least two points on the horizontal branch, except for the most massive of stars, which have extremely rapid evolution. We slightly increased the number of isochrone ages used in the grid to 616 from 498 in SPYGLASS-I, but left the mass and metallicity grids unchanged. We interpolated values for $G$, $G_{BP}$, and $G_{RP}$ off the resulting grid according to the values of age, mass, and metallicity for each model star, producing photometry for each one.

Finally, we merged the photometry of unresolved binaries. In SPYGLASS-I, all model stars were taken to be at the same distance due to the survey's relatively limited 333 pc distance horizon, which reduced the differences in the unresolved rate across the sample (see the discussion in SPYGLASS-I). However, with our expansion of this search to 1 kpc, a more versatile approach is required to properly capture the contribution from unresolved binaries at these larger distances. To do this, we split the selection of Gaia stars into 8000 bins sorted by distance, all but one being equally populated with 6621 stars. For the distance to the stars in each bin, rounded to the nearest 1 pc, we generated a model with 10 million sample stars. At that distance, we computed angular separations for each generated linear separation, merging the photometry of all stars with separations below 1 arcsecond, following SPYGLASS-I. We did not generate multiple models for bins with the same rounded distance, so in practice we generated models at 1 pc increments for bins with distances above $\sim$90 pc. The spacing between bins was increasingly wide closer to the sun, with the two closest bins to the sun having average distances of 19 and 30 pc. Coarser distance sensitivity within 90 pc is however of limited concern for youth assessment, as SPYGLASS-I showed that our sensitivity to young groups in that region is relatively weak due to dominant geometric projection effects. The total number of models generated was 939. This step completed our set of intrinsic models, which were later modulated with reddening and distance to produce model apparent magnitudes for individual stars. 

\begin{figure}
\centering
\includegraphics[width=8cm]{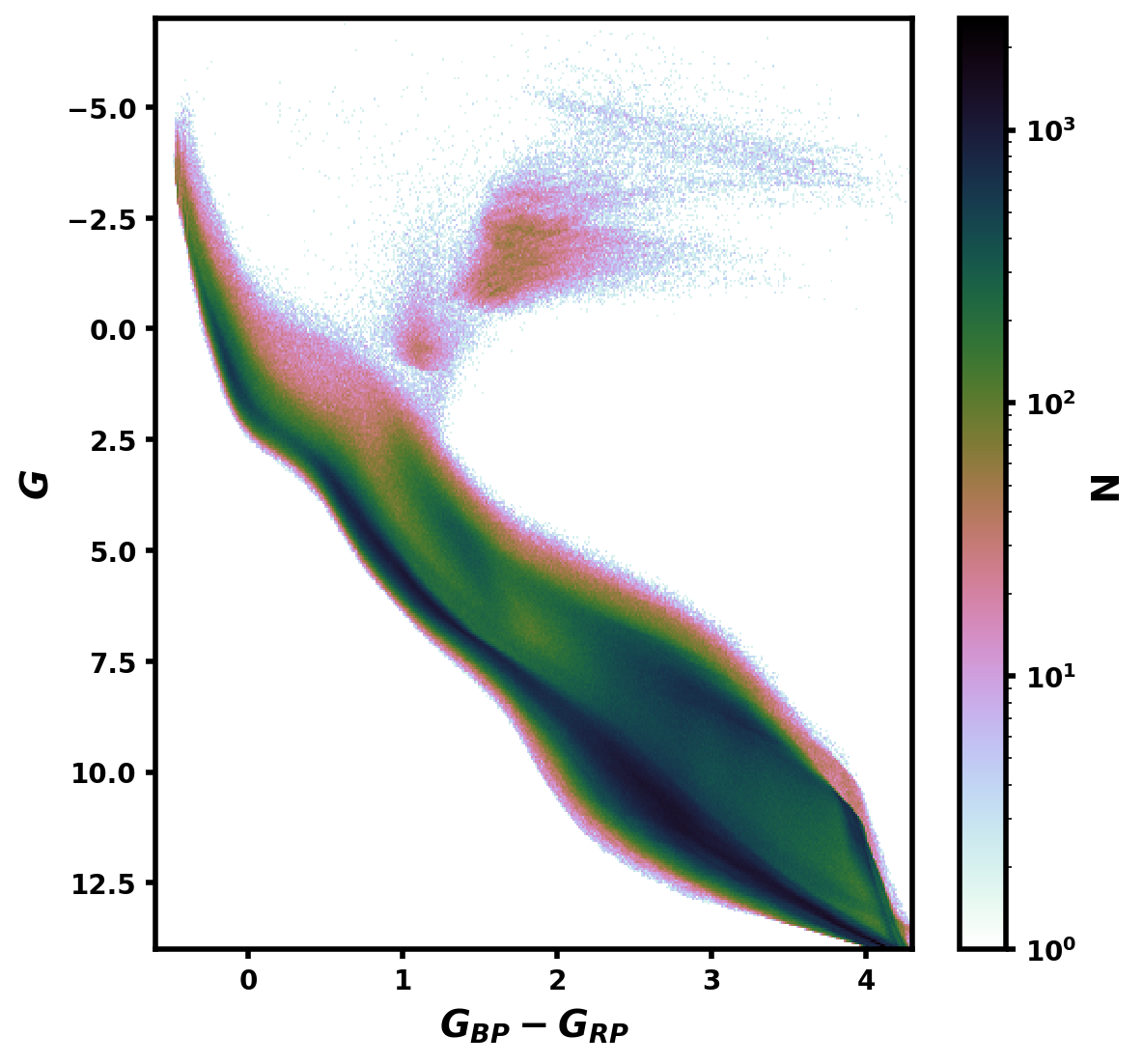}\hfill
\caption{An HR diagram demonstrating our sample model using the new DR3 photometric system. This example uses a distance of 500 pc for merging binaries, although we also generate 938 other models for comparison with stars at different distances. Like in SPYGLASS-I, there are some streak-like density anomalies on the giant branch and far upper main sequence. These are caused by the speed of stellar evolution in these regions, however youth is typically not contentious there, so sampling density issues are unlikely to be important.}
\label{fig:model}
\end{figure}

A sample model at 500 pc is provided in Figure \ref{fig:model}. While the updates to the Gaia photometric system and binary management were important for the generation of statistics, they were also subtle to the eye, making this model and the others we generated difficult to distinguish from the SPYGLASS-I models. There was still some slight under-sampling that produced artefacts on the subgiant branch and OB sequence like in SPYGLASS-I, however our finer grid sampling in these models reduced these effects compared to SPYGLASS-I. The effects of these artefacts nonetheless have little effect on our statistics, especially given that less massive stars dominate our young sample. 

\subsection{Generating Star Statistics} \label{sec:starstats}

In SPYGLASS-I, distances to Gaia stars were directly combined with apparent magnitudes to generate absolute magnitudes, with uncertainties derived from the photometric and parallax uncertainties. While this produces an accurate representation of the absolute magnitude uncertainty for an individual magnitude, the Gaia $G$, $G_{BP}$, and $G_{RP}$ absolute magnitudes are all covariant in distance, and as a result, treating all three as independent provides a imperfect statistical representation of these uncertainties. To resolve this, we instead introduced the distance corrections directly into the model. We randomly generated a set of distances for each star from a gaussian centered on the mean \citet{BailerJones21} geometric distance with high and low 1-sigma intervals equal to the upper and lower limits of the corresponding distance measurement. These became model distances, which, when combined with the model magnitudes, generated model absolute magnitudes, which can be directly compared to our Gaia observables. 

This is the same approach used to introduce reddening, both in SPYGLASS-I and in this paper. We interpolated reddening values from the \citet{Lallement19} reddening maps, along with the upper and lower uncertainty intervals. We then generated a set of model reddenings drawn from the reddening probability distribution for each star in the model. These reddenings were applied to the model stars like for distance, changing the model accordingly. The direct addition of model reddening and distance values to the intrinsic stellar models therefore produced a model specific to each star, which fully accounts for the covariance of the apparent magnitudes in reddening and distance. 

While this approach makes our uncertainty handling more accurate, photometric uncertainties from Gaia are often much smaller than those induced through distance measurements, and the choice to build the distance uncertainty into the model does not improve the density of model stars. As a result, statistical measurements which only consider photometric uncertainties result in many stars with tightly-constrained Gaia photometry having such tight photometric distributions that few, if any model stars lie within a few sigma in magnitude-space. We retrospectively identified a similar issue on rare occasions in SPYGLASS-I, in which a small number of very low-uncertainty stars were excluded from our sample through the requirement that at least 50 model stars must reside within 1-sigma of each Gaia star in magnitude space. This choice was intended to remove white dwarfs and other stars not covered within our model, however it inadvertently also excluded a few high-quality candidates. Modifications to our methods are therefore necessary to maximize the number of good comparisons to model stars for each Gaia star. 

We therefore introduced factors which account for internal uncertainties in the models themselves. These model uncertainties were separated into two components: the uncertainty within the model itself, and the coarseness of our model population in sparsely-populated sections of the CMD. Two new uncertainty factors were therefore introduced into the formula for the probability of a sample star being consistent with a Gaia observation, updating the equation presented in SPYGLASS-I to the following:
\begin{equation} \label{eqn:pyoung}
p(y|g) \propto p(y)p(g|y) = \prod_{i}exp\left(-\frac{(g_i-x_i)^2}{2(\sigma_{g, i}^2+\sigma_{x, i}^2+\sigma_{s,i}^2))}\right)
\end{equation}
In this framework the subscript $i$ multiplies over each observable in x, with $g_i$ being the observables of the Gaia star, $x_i$ being the observables of the model star, and $\sigma_{g,i}$ being the uncertainty in Gaia observables. The only prior $p(y)$ we have accounts for the oversampling of young stars in age space, and the corrective factor is equal to the age of the star. 

The last two $\sigma$ values in the denominator of the exponential are therefore new to this publication and not included in SPYGLASS-I. The first new uncertainty, which we call the model uncertainty, $\sigma_{x, i}^2$, is set equal to 0.02 mag. This value is chosen to represent the approximate minimum magnitude difference induced by a 0.05 dex change to metallicity, with that definition meant to represent the magnitude change induced by a minimally resolvable change in the model \citep[][]{PARSECChen15, GarciaPerez16}. This 0.02 mag value is also considerably smaller than the width of typical main sequences of star clusters, ensuring that it does not introduce any additional uncertainty not already seen in populations with presumed identical properties. This additional uncertainty factor has little effect for stars with uncertain Gaia photometry, however it ensures that stars with extremely well-constrained Gaia photometry are still able to take advantage of our model sampling density.

Most anomalies produced by small Gaia uncertainties were resolved by the addition of this model uncertainty parameter. However, regions in the CMD with particularly low model densities still occasionally produced anomalous youth results. These issues are most evident in the region where the subgiant branch intersects the pre-main sequence, where our logarithmic age sampling populates this region well with young stars, but the sampling of subgiants remains sparse. As such, stars identified as young in this region typically had few or even no model subgiants within 1$\sigma$, despite the presence of subgiants in nearby sections of the CMD. There was also a region below the main sequence where a small group of stars sits below the range of our model populations, most likely due to the presence of relatively rare stars with metallicities outside the range our model considers. Both of these regions occasionally see stars identified as young, despite the fact that their position casts significant doubt on their youth. 

The need to ensure that these populations have sufficient model coverage for a meaningful youth assessment prompts us to add the second new uncertainty factor to the denominator of the exponent in Equation \ref{eqn:pyoung}, $\sigma_{s,i}^2$. We set this value equal to 0.1 magnitudes multiplied by a new integer $j$. We found that stars with questionable youth assessments generally had $N < 200$ model stars within 1$\sigma$ in magnitude space, so we use the uncertainty term 0.1$j$ to account for coarseness of the model grid, where $j$ is initialized to 0 and increased by 1 until $N > 200$. While this does have the effect of somewhat blurring our model beyond its resolution for stars with $j>0$, the base uncertainty addition of 0.1 magnitudes is still capable resolving age differences as small as 5 Myr on the pre-main sequence, a level of resolution exceeding what is required for a broader youth determination. For white dwarfs, these uncertainty increases required to get $N > 200$ can be unphysically large due to our lack of white dwarfs in our modelling, although given that these stars are not of interest to our work, poor handling of them is not a concern, and no white dwarfs end up in our young star sample. 

\subsection{Selecting Young Stars} \label{sec:ysselect}

Our search for young stars was performed on the quality-restricted sample of 53 million stars described in Section \ref{sec:data}.  However, due to the changes to our stellar probability analysis in Section \ref{sec:starstats}, we found that the $P_{Age<50 Myr}>0.1$ cut marked stars as young that are up to 0.5 magnitudes lower on the pre-main sequence compared to SPYGLASS-I. Most of the changes to $P_{Age<50 Myr}$ appear to result from our new Bayesian management of distance and Gaia DR3 improvements to distance uncertainty, which together significantly reduce uncertainties for the absolute magnitudes in each Gaia color band. Since young stars are rare relative to older stars, poorly-constrained measurements tend to default to a low $P_{Age<50 Myr}$. The result of improving our uncertainties therefore widens the range of stars with $P_{Age<50 Myr}>0.1$, which in the clustering stage merges populations that perhaps should not be merged. 

We therefore changed the selection limit to $P_{Age<50 Myr}>0.2$, which brings the parameter space identified as young roughly in line with SPYGLASS-I while improving the contrast between structures and the background. Given our calculated probabilities, that restriction is expected to produce a sample where approximately two-thirds of the total stars have Age$<50$ Myr. This choice therefore accepts slightly elevated contamination in exchange for the inclusion of populations near the upper end of our age limit where the field binary sequence and pre-main sequence overlap the most. This choice to priorize these older populations over contamination synergizes well with the clustering method we use to identify young populations, which is built to identify overdensities in the presence of a significant background. Most of the non-young model stars that contribute to $P_{Age<50 Myr}$ for values near 0.2 are also binaries, so they can be suppressed using RUWE cuts, which we impose in Section \ref{sec:clustering}. The final population of photometrically young stars after updating the $P_{Age<50 Myr}$ threshhold is 418611.


\section{Young Populations} \label{sec:populations}

\subsection{Clustering} \label{sec:clustering}

Our clustering in this publication uses HDBSCAN \citep{McInnes2017}\footnote{further information on HDBSCAN can be found at \url{https://hdbscan.readthedocs.io/en/latest/how_hdbscan_works.html}}, and almost entirely follows the choices made in SPYGLASS-I. We cluster in five-dimensional $(X,Y,Z,c*v_{T,l},c*v_{T,b})$ space, using the constant $c$ to equalize the typical scales of the space and velocity components, which is set to $c=6$ pc km$^{-1}$ s. We set the HDBSCAN parameters {\tt min$\_$samples} and {\tt min$\_$cluster$\_$size} to 10, and set $\epsilon$\footnote{which is represented by the HDBSCAN clustering parameter {\tt cluster$\_$selection$\_$epsilon}} to 25, with the latter being the parameter used in SPYGLASS-I to allow for the merging of groups with similar enough distributions to hint to mutual connections. We only used excess of mass (EOM) clustering, which identifies groups by their persistence across clustering scales. Leaf clustering, which identifies the smallest scales of overdensities useful in analyzing substructure, is beyond the scope of this publication. Starting with the sample of candidate young stars drawn from our quality-restricted sample, we required that $d/\sigma_d > 25$ to ensure manageable distance spreads in galactic coordinates, following SPYGLASS-I. This reduces the sample size to 199434, making this restriction more impactful in this publication than it was in SPYGLASS-I. This cut is nonetheless worthwhile to improve clustering for the closer and more accessible populations that are our focus. Finally, we added a new cut requiring RUWE $<1.4$. This is the loosest of the typically-accepted RUWE cuts for removing binaries, and we find that this cut is useful to reduce the background of misidentified older binaries \citep[e.g.,][]{Bryson20,Stassun21}. The final sample of high-quality young stars contains 181524 stars after these restrictions. Once these cuts were made, we applied HDBSCAN to identify clusters, resulting in 228 populations being identified within 1 kpc, containing about 39000 photometrically young stars. 

One additional and particularly important change was made to the clustering methodology for this paper: the removal of the cut on the cluster persistence factor. In SPYGLASS-I, this factor was used to remove clusters associated with reddening anomalies, however in doing so it likely removed many real features. By removing this cut, we reintroduced many new and potentially interesting features, however we also greatly increased the potential influence of anomalous features. Our larger search distance exacerbates this issue by introducing significant new reddening anomalies through both the addition of new dense clouds into the search area and through the addition of more area within the backgrounds of these clouds, which can result in reddening anomalies spanning hundreds of parsecs. As a result, the remainder of this section primarily concerns cluster validation, including both the removal of likely spurious features, and the revision of the cluster member lists. 

\subsection{Reintroduction of Photometrically Ambiguous Populations} \label{sec:extpops}

Our clustering methods identified many young populations, both known and unknown. However, since we only clustered on a subset containing high-quality and photometrically young stars, there are certain to be other members that are not included in these base samples. The generation of complete group populations therefore requires the reintroduction of stars that are co-spatial with identified members in space-velocity coordinates, but are not photometrically identifiable as young. To do this, we followed the same method employed in SPYGLASS-I, defining a distance metric equal to the distance to the 10th nearest young member ($d_{10}$), which is similar to the metric used in HDBSCAN clustering. Candidate members were subsequently identified as having a $d_{10}$ smaller than the largest value for an identified young member. To preserve information on the credibility of each candidate member, we computed clustering proximity ($D$, formerly ``strength'' in SPYGLASS-I), which is a measure of a star's centrality within the space-velocity distribution of a group. It is defined such that the star in the cluster with the largest $d_{10}$ has a value of zero, and the star with the smallest $d_{10}$ (i.e., the most central member) has a value of 1, with a linear scale in between. We later use $D$ as the basis for computing cluster membership probabilities, $P_{mem}$, in Section \ref{sec:dpmem}.

Our search for extended candidate populations was applied to the nearly unrestricted set of 94 million stars discussed in Section \ref{sec:data}. This full stellar sample contains all stars for which membership and the feasibility of follow-up observations can be assessed, and provides maximally complete populations for the groups we identified. We used the subset of this sample for which we generated youth statistics in validation work, as the properties of a group's extended population of candidates strongly inform its coherence (see Section \ref{sec:vetting}). In total, approximately 3 million stars were identified as candidate members of a young association. The nearly 100-fold increase in the membership of these expanded populations relative to the 39000 young stars used to identify them suggests significant field contamination in many of these populations, which we address in Section \ref{sec:pmem}.


\subsection{Cluster Vetting} \label{sec:vetting}

\begin{figure*}
\centering
\includegraphics[width=18cm]{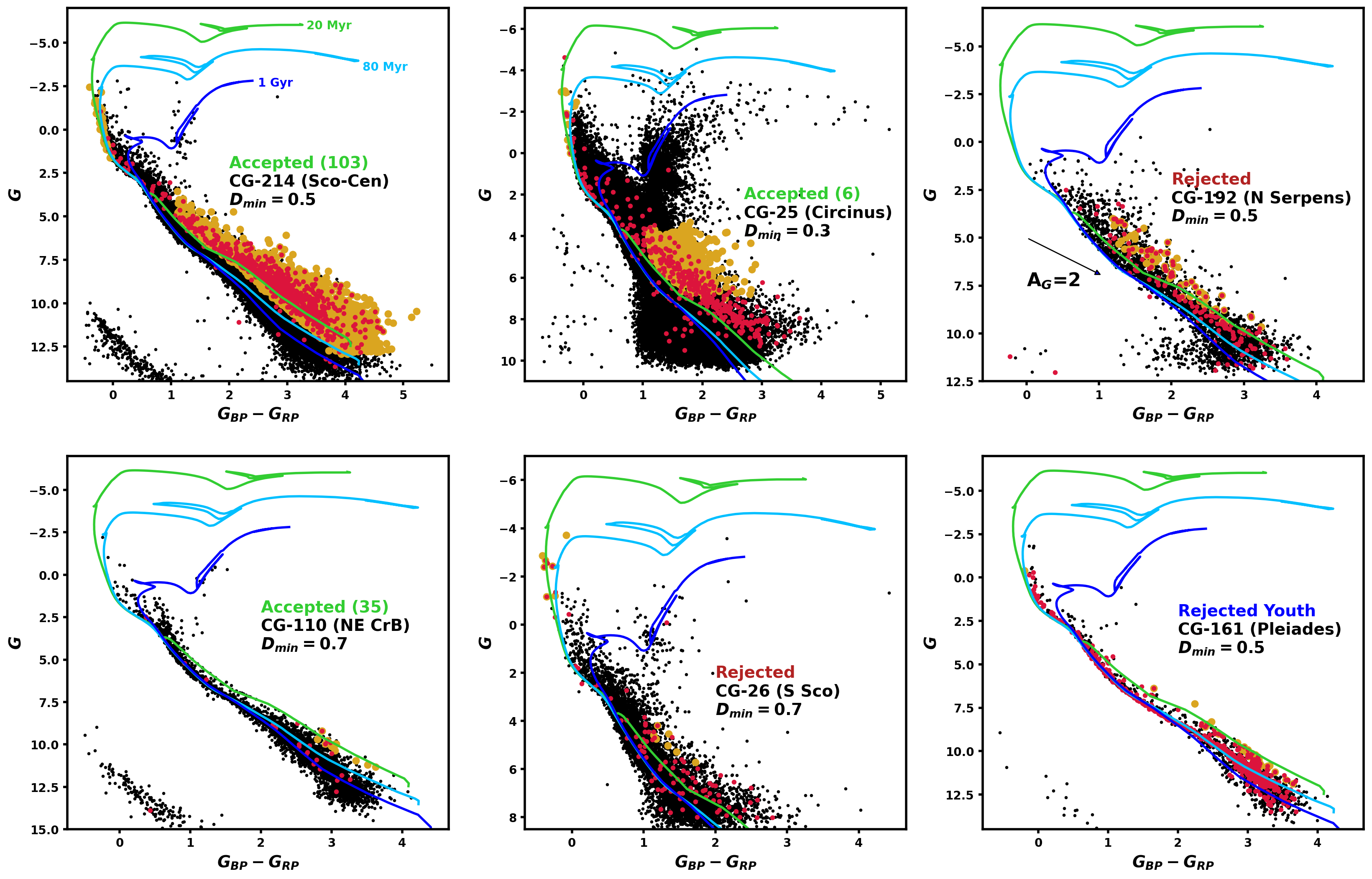}\hfill
\caption{Example groups which demonstrate our vetting choices. For each group, we show the photometrically young sample used to identify the group (large yellow dots), the full quality-restricted sample of candidate members which includes photometrically older candidates (small black dots), and a $D$-restricted subset of these candidates limited to stars near the cluster center (red dots). The value of $D_{min}$ used for restriction is annotated, along with the candidate group ID and location. For accepted groups, the assigned SCYA IDs are provided in brackets. 1 Gyr, 80 Myr, and 20 Myr isochrones are provided for reference, which represent the field population, maximum age of SPYGLASS populations, and typical young groups, respectively. The top-left panel shows the Sco-Cen association, where restriction in $D$ shows a strong sequence which overlaps with the photometrically young sample. The top-middle panel shows the Circinus complex, which is real but very distant, resulting in most young stars being found just below the subgiant branch. While this typically indicates a spurious group, OB stars are present and the $D$-restricted subset reveals a strong sequence, so we accept this population. The top-right population is also identified by the region below the subgiant branch, however there are no OB stars, the $D$-restricted main sequence is mostly focused on the field sequence, and that $D$-restricted sequence appears consistent with a reddened field sequence (the reddening vector is shown for reference). We therefore reject it. The bottom-left panel shows a new and very tenuous population. Despite the small population of young stars in the $D$-restricted sequence, much of the young sequence originally identified is still there, while the field population is comparatively nearly gone. We therefore accept this group. The bottom-middle panel shows a likely false group in southern Scorpius. While this sample has OB stars, it is otherwise identified by the region beneath the subgiant branch, and the $D$-restricted sample does not skew towards a young sequence, instead looking like a depopulated version of the full sequence. We therefore reject it. The final (bottom-right) panel shows the Pleiades, an older population that straddles the 80 Myr isochrone. This isochrone marks our limit for defining groups as young, so this group and any older than it are rejected from the young sample, and handled in Appendix \ref{app:old}.}
\label{fig:clustervetting}
\end{figure*}


To clean the sample of false clusters produced by our inclusive clustering result, we must identify indicators that are associated with the false groups we often see. Many metrics for cluster vetting exist, however they can also be difficult to quantify and apply uniformly due to the highly varied positions, shapes, and environments of the young populations identified. As a result, we found that visual input from a human was often required for proper assessment. In this section we define the indicators used to identify false positives, and vet these populations by hand in accordance with those metrics. 

Our first major indicator of false positives arises from poorly-corrected reddening, which results from the spatially-coherent errors of the \citet{Lallement19} reddening maps. This is by far the most common cause of false cluster identification, especially in the most heavily reddened environments. These reddening anomalies often result in large swaths of old stars behind a cloud being identified as young. However, since the reddening vector moves stars nearly parallel to the main sequence, field stars that are falsely identified as young due to uncorrected reddening are generally limited to those that were originally on the subgiant branch, but were reddened towards the lower-right on the HR diagram, placing them on the pre-main-sequence. Moderate reddening of subgiants yields colors of $0.5<G_{BP}-G_{RP}<1.5$, defining a region on the HR diagram which usually contributes minimally to young stellar populations (with approximate masses $0.75 M_{\odot} < M < 1.4 M_{\odot}$), but can become the dominant region occupied by photometrically young objects if reddening is improperly corrected. Most well-known young groups, such as Sco-Cen, Orion, and Vela have populations of photometrically young stars with between 1\% and 5\% of their membership in this reddening error-contaminated region, and this low fraction is expected for a typical IMF \citep[e.g.,][]{Chabrier05}, as the less massive K and M stars are much more common compared to the earlier-type stars in this color range. Sco-Cen, being largely gas-free throughout most of its extent, has a fraction of 1.2\%, while Orion, which has more gas present, has a higher value, at 2.7\%, indicating the introduction of minor reddening anomalies. More consistently reddened environments and those at larger distances frequently had much higher values, however we found that for all but the most distant associations, spurious populations were reliably identified by fractions exceeding 50\%. 

The most distant regions are however exceptions, as with the quality restrictions we employ prior to clustering, populations with $d \ga 800$ pc often have limited representation beyond this often-contaminated region below the subgiant branch. Assessments of these distant populations must therefore include the extended populations, which are not restricted on distance uncertainty and therefore extend to dimmer magnitudes than the population used for clustering. A real cluster among these distant populations can generally be identified by the presence of a sequence extending parallel to an isochrone and directly overlapping with the stars just below the subgiant branch used to define the group. 

The other major identifier we use to detect false positives in clustering is based on a lack of internal coherence. In real young populations, the space and velocity coordinates should both show a concentration of genuine members, in which credibly young stars are more common closer to the group center, and field stars become increasingly dominant further out. False groupings tend to have either no central concentration of young stars, or concentrations guided by reddening patterns, in which more central stars, specifically in spatial coordinates, may be on a different sequence produced purely by the locally heavy reddening of a field sequence. We assessed the central concentration of the young sequences using the quality-restricted extended populations of candidates, which are drawn from the age-unrestricted population of 53 million stars for which we have measurements of youth probability. We restricted these populations in $D$, observing whether a young sequence emerges more clearly with restricted $D$, or whether the sequence could be explained as just a random assortment of field stars. 

The metrics we have outlined, which focus on central concentration and subgiant branch reddening contamination, are used to assess whether each association in our catalog is real or a spurious detection. We removed clusters that failed either of these metrics, making the final judgement for each association by hand. In Figure \ref{fig:clustervetting}, we provide some illustrative vetting examples, which cover the full range of marginal cases which arise in our set of candidate populations. 

The top-left panel shows the population of Sco-Cen as a prototypical example of a real group \citep{Preibisch08,Pecaut16}. The young stars used to identify it cover much of the pre-main sequence, and low-mass stars are much more numerous among them, as the Initial Mass Function would predict \citep[e.g.,][]{Salpeter55,Chabrier05}. There is also no clear concentration of stars below the subgiant branch (1.2\% with $0.5<G_{BP}-G_{RP}<1.5$), suggesting that reddening has a minimal effect on young star identification here. Restricting the extended population in $D$ results in an increasingly clean stellar sample with increasing $D$, suggesting that while there is field contamination near the edges, the center is very pure. Most of the stars on the $D$-restricted sequence reside above the 20 Myr PARSEC isochrone \citet{PARSECChen15}, meaning that it is clearly within our target age range. The remaining five are all much less obvious in their vetting decision due to deficiencies in one or more of the observables that makes Sco-Cen's sequence so convincing. Those examples therefore cover all major choices involved in group vetting. 

The top-center and top-right panels of Figure \ref{fig:clustervetting} both show populations mainly identified through stars in the reddening-vulnerable region below the subgiant branch. The top-center panel shows the Circinus complex, which is a known young population \citep[e.g.,][]{Reipurth08Cir}, while the top-right panel shows a reddening anomaly behind the Serpens complex, which we label Candidate Group (CG) 192. There are a few differences between them. First is the OB stars, which are present for Circinus, but entirely absent for CG-192. While smaller populations will not always have OB stars, their presence is generally a strong indication that a real population is present. However, the much more telling vetting indicator in this case is the sequence of the extended population. In Circinus, restricting the population in $D$ reveals a strong pre-main sequence well above the 20 Myr isochrone, following a standard isochrone track that passes through the photometrically young founding sample. The young founding sample for CG-192 is also elevated above the 20 Myr sequence, which would predict the presence of even more stars along this sequence at lower masses, like in Circinus. Instead, the sequence curves down across the isochrones, settling along the right edge of the field main sequence. Rather than being a young sequence, this appears consistent with a reddened copy of the field sequence, pushed to the lower-right of the CMD by approximately 2 magnitudes of reddening. With its location near the center of the Serpens Complex and adjoining dense clouds, an anomalous population here is not unexpected. 

The bottom-left panel of Figure \ref{fig:clustervetting} shows CG-110, which is a lightly-populated and tenuous association with a clear central concentration. Its extended population is dominated by the field, however with $D>0.7$, stars on the young sequence approach the field in abundance, with about one-third of stars being found near the 20 Myr isochrone rather than the field sequence. Furthermore, none of the young stars used to define CG-110 are near the subgiant branch, so there is no indication of reddening interference. The result is a population that, while very small, satisfies every requirement of a genuine population, resulting in it being accepted. 

The bottom-center group (CG-26), however, is less convincing. It has a few OB stars, which would ordinarily be predictive of a substantial young population given their rarity, however restricting in $D$ reveals a sequence that looks like a depopulated version of the field sequence, without any young population emerging. While it is tempting to define a population based only on the OB stars, groups that we know of have a range in masses, so we must require that credible evidence of a young sequence appears over the entire region where it should be visible. In this particular case, it is unclear why the OB stars are present, however this is in a crowded region of the central Milky Way in southern Scorpius, so it is possible that the group was identified mainly by reddening anomalies. Groups from reddening anomalies tend to have broader velocity spreads, so the area connected to it may have included some genuinely young stars on its periphery, making its youth appear more credible. Nonetheless, this is not a group that produces internally-consistent results, so we remove it.  

Finally, in the bottom-right panel of Figure \ref{fig:clustervetting}, we show the Pleiades, which is an example of a group which is real, but is not especially young. These older groups were occasionally identified as young populations, typically through some combination of reddening and a strong binary sequence. We used the 80 Myr PARSEC isochrone to set that limit, which is slightly younger than the typically-accepted ages for the Pleiades (100--160 Myr; e.g., \citealt[][]{Stauffer98,Gossage18}). Anything found with a pre-main sequence below the 80 Myr isochrone is excluded from further analysis, but we do report their populations in Appendix \ref{app:old}.

\begin{figure*}
\centering
\includegraphics[width=16cm]{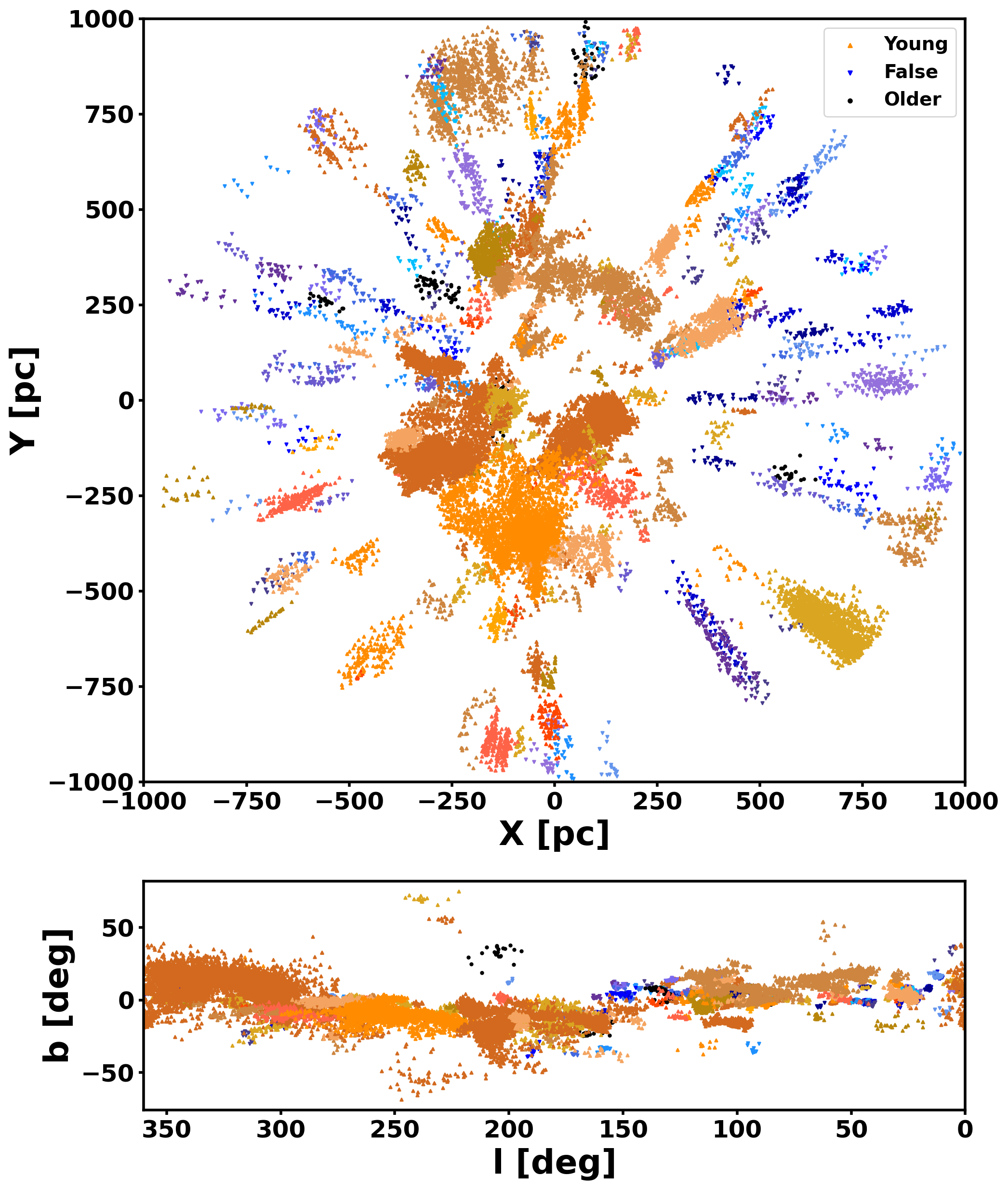}\hfill
\caption{Clusters identified by HDBSCAN, shown in XY galactic cartesian coordinates and l/b galactic sky coordinates marked by their vetting status. We introduce small differences in hue to distinguish overlapping groups, but we broadly use orange-ish shades marked by up-arrows represent real groups, while blueish shades marked by down-arrows represent false groups. Real but older groups are shown as black circles. The extensive rays of false groups towards the right side of the XY plot are mainly produced by reddening anomalies in the foregrounds in Serpens, Aquila, and the Pipe Nebula. }
\label{fig:rvfgroups_xylb}
\end{figure*}

We show all groups in the sample in Figure \ref{fig:rvfgroups_xylb}, colored by whether they are accepted, rejected, or real but old. Groups that are accepted by our vetting process are treated as genuine young associations for the remainder of this publication. We give all such groups new indices within a new ``SPYGLASS Catalog of Young Associations'' (SCYA), which are used to refer to them later in this paper. A total of 116 groups pass vetting, and are given a SCYA ID as a result. 

The clusters removed are often grouped in sky coordinates, occasionally having essentially identical spatial distributions (see Figure \ref{fig:rvfgroups_xylb}). This tends to provide reliable verification for a group's removal, as the radial stacking of groups is almost always produced by reddening anomalies that result in the systematic misidentification of stars as young in their backgrounds. The Serpens and Cepheus clouds produce some of the most frequent false detections, but most notable molecular clouds in the solar neighborhood produce at least one anomalous group behind them. However, these tend to be easily removed using the already-imposed vetting methods. While we did consult reddening maps when making our final vetting choices to confirm reddening suspicions, no groups were removed solely due to the presence of a reddening anomaly, as it is common for genuine populations to be embedded in heavy reddening. 

\section{Membership Probabilities} \label{sec:pmem}

In previous SPYGLASS papers, the clustering proximity parameter $D$ (which includes both spatial and kinematic distance) was used as a proxy for a star's likelihood of membership, with stars more central in their parent distribution being deemed more probable members than stars near the boundary with the field. However, since that value only calculates distance to young neighbors, it ignores the local density of the field population. As a result, any given value of $D$ could have very different implications depending on the density profile of both the group and the field. A true membership probability ($P_{mem}$) would compare the relative sizes of the field and the young population in the vicinity of the star, similar to the statistical approach used by \citet{Sanders71} and numerous others. We therefore created an algorithm to produce conversion maps between $D$ and $P_{mem}$ for each population in our sample. Results are provided for each group that passes vetting. 

\subsection{Corrective Factors for Stellar Populations} \label{sec:popcf}

\begin{figure}
\centering
\includegraphics[width=8cm]{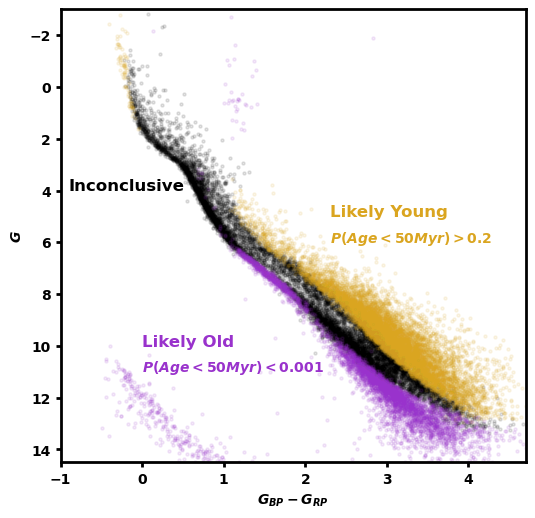}\hfill
\caption{Figure showing the regions of the HR diagram that we identified as reliably old ($P_{Age<50 Myr} < 0.001$) or young ($P_{Age<50 Myr} > 0.2$), using Sco-Cen as an example. Young stars are identified on the pre-main sequence and the tip of the OB sequence, and old stars are confidently identified on the lower main sequence, giant branch, and white dwarf cooling sequence. The faintness of most stars in those reliably young or old sets makes the fraction of the respective populations that they represent a strong function of distance once faint stars begin to become invisible to Gaia.}
\label{fig:youngoldsequences}
\end{figure}

We must first establish a way of reliably estimating the populations of the young group and the field separately so that they can be compared. Using the quality-restricted extended candidate samples, we generated near-certain populations of group and field populations using their youth probability. We defined stars in the quality-restricted extended populations with $P_{Age<50 Myr} > 0.2$ as young (following Section \ref{sec:ysselect}), and stars with $P_{Age<50 Myr} < 0.001$ as old. The locations of these identified sequences are shown in Figure \ref{fig:youngoldsequences}, using Sco-Cen as an example. Identified young stars were, as expected, pre-main sequence and OB stars, while reliably old stars were below the pre-main sequence, on the giant branch, and on the white dwarf cooling sequence. With these confident populations established, total populations of the young and old samples could then be estimated by calculating the fraction of stars missed by the young and old selections. 

\begin{figure}
\centering
\includegraphics[width=8cm]{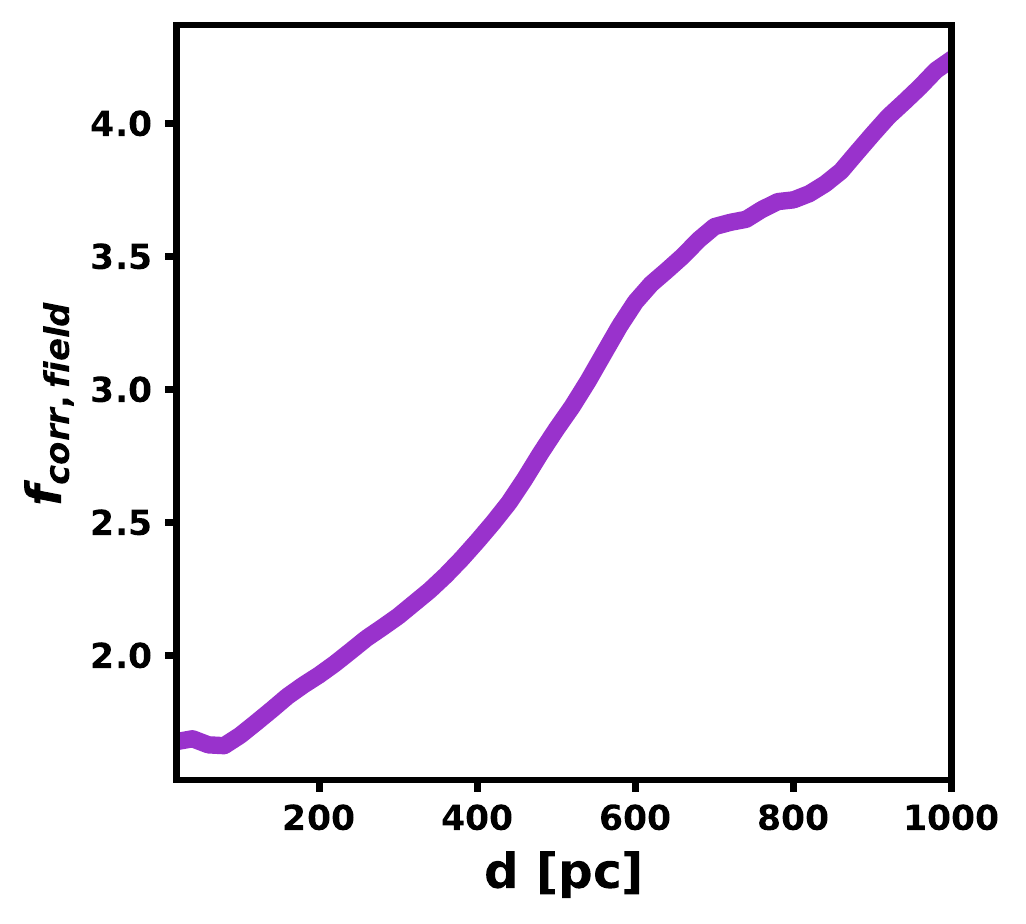}\hfill
\caption{The curve for the field population corrective factor, which is the inverse of the fraction of total field stars identified as old using $P_{Age<50 Myr} < 0.001$ (see Fig. \ref{fig:youngoldsequences}). The increase with distance reflects the less reliable youth assessment for stars that are higher on the main sequence.}
\label{fig:fieldcorrsequence}
\end{figure}

The field stars are relatively straightforward for which to compute missing fractions, as the field tends to have similar photometric properties regardless of the direction, especially within the galactic plane where most associations reside and metallicity gradients are minor. Our sensitivity to stars is distance-dependent, so we computed the correction for field stars as a function of distance. To do, this we took the stars in our Gaia sample that are not likely young ($P_{Age<50 Myr} < 0.2$), and computed the fraction of those which have $P_{Age<50 Myr} < 0.001$ across 20 bins from 0 to 1 kpc. After smoothing the results with a Gaussian kernel, we had a curve for the field abundance conversion as a function of distance. We provide this curve in Figure \ref{fig:fieldcorrsequence}. For each star with $P_{Age<50 Myr} < 0.001$ at a given distance, a corresponding corrective factor can be read off, providing an expected number of missing stars for each identified field star. 

We then estimated the missing fraction for the members of the young population. As shown through the completeness discussion in SPYGLASS-I, SPYGLASS recovery rates are age-dependent. However, like for the field population, the much larger search horizon employed in this work prevents us from reaching the bottom of the main sequence in some cases, resulting in distance also affecting completeness. Unlike those nearby groups used to calculate completeness in SPYGLASS-I, there is not nearly enough coverage in distant groups to reliably factor distance into that sort of analysis. We therefore developed a conversion based on the populations available to us in each group. We plan on addressing the issue of completeness in more detail in an upcoming paper, which will provide a statistical view of the populations identified in this paper \citep{kerrspyglassvii}. 

Our missing fractions for the field allowed us to estimate the total field population in each group. The population of the young group itself should then be the difference between the population of all candidates and the field population. The corrective factor which provides the total group population from the young sample is then computed as the size of the young group population divided by the number of stars with $P_{Age<50 Myr} > 0.2$.  While field stars often dominate candidate populations, occasionally overwhelming the contribution from the young association, our vetting choices ensure that for a sufficiently restricted range of $D$, the population of the young sequence approaches or exceeds that of the field, providing robustly measurable young stellar populations which enable this calculation.

\begin{figure}
\centering
\includegraphics[width=8.5cm]{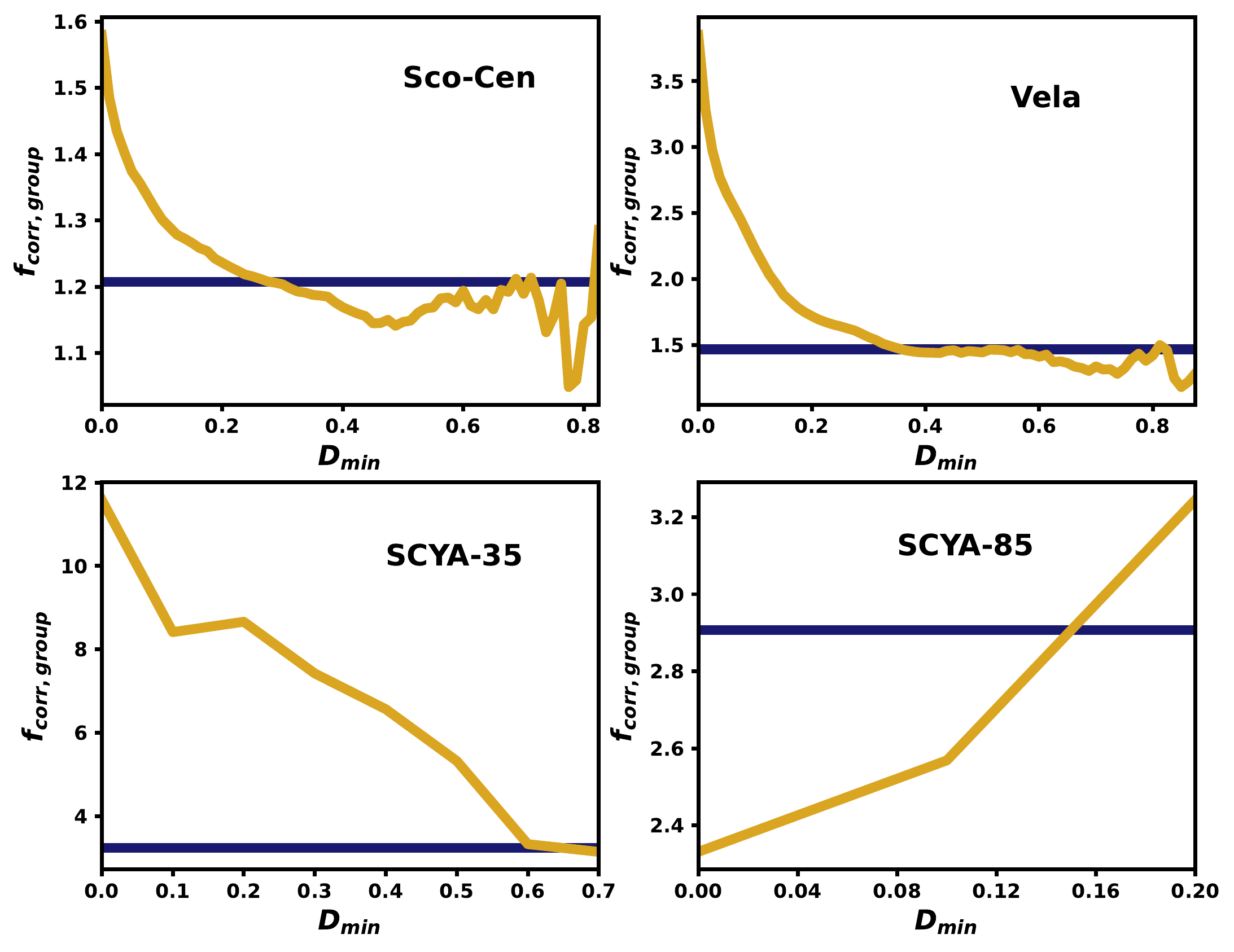}\hfill
\caption{Curves of the corrective factors to the populations of young groups, plotted in yellow, as a function of $D_{min}$, the minimum value of $D$ that the population is restricted to for a given point. The final corrective factor is shown as the dark horizontal line. Sco-Cen and Vela are both prototypical examples where the final value of $f_{corr,group}$ is selected as the peak of the corresponding KDE. SCYA-35 (CG-110 pre-vetting) plateaus only for very restrictive choices of $D_{min}$ and SCYA-85 has only three corrective factor measurements, so in these cases we calculate the final $f_{corr,group}$ as the mean of the last two values, which tend to be most dominated by group members. }
\label{fig:ygroupcorr}
\end{figure}

For large and kinematically distinctive groups, this method produced consistent conversion factors for our young populations, across a range of choices for $D$ restriction. However, for many smaller groups, the sample of young stars was small enough relative to the field that this calculation was dominated by Poisson uncertainty in the population of field stars, skewing it. We nonetheless found that by restricting the minimum value of $D$, the corrective factor changes as the group becomes increasingly dominant over the field, until it plateaus at the presumed corrective factor for the group. The $D$ values we used to restrict the selection are the same as those later used in Section \ref{sec:dpmem} to calculate $P(mem)$ as a function of $D$. We only ran these calculations for stars above a certain $D$ when at least 5 objects were likely old or young stars, which are defined as having $P_{Age<50 Myr} > 0.2$ or $P_{Age<50 Myr} < 0.001$. This ensures that there is a population of stars firmly in either the field or the group population which help set these fractions. Upon calculating the corrective factor for a range of $D$ restrictions, we produced a KDE of the resulting corrective factors, and computed the final correction value as the peak of that KDE. Some examples of the $D_{min}$ vs $f_{corr,group}$ curve are shown in Figure \ref{fig:ygroupcorr}, alongside the fit from the KDE. In most cases the KDE peak returns the plateau in the corrective factor achieved once the group population dominates over the uncertainty in the field calculation. These plateaus are occasionally accompanied by some variation caused by internal age differences as a function of $D$ which the KDE averages over. However, there are also a small number of groups with too few likely old or young stars to generate calculations over a wide range of $D$, and 3 additional extremely tenuous groups which are dominated by the field for most $D$ restrictions, making their plateau undetectable to the KDE. In these cases, we computed the corrective factor as the average of the two solutions with the most restricted $D$, and in all cases this produced a corrective factor within the range expected by visual inspection. 

\subsection{Producing a $D$-P$_{mem}$ Curve} \label{sec:dpmem}

The reliably young and old populations, combined with the corrective factors that account for the missing fraction, allows us to compute the fractional abundance of members as a function of $D$. For binning, we chose 10 bins covering $0<D<1$ as a base selection, and for more populated groups we doubled the number of bins until less than 200 stars exist in the bin just above $D=0.4$. More populated groups often dominate the center of their parameter space and only lose dominance near the edges, and more bins allows that dropoff to be resolved. $D=0.4$ is an intermediate value that avoids the occasional field domination of low $D$ and the inconsistent density at high $D$, hence why it is a good place to assess the need for adjusting bin density. We also required that there be at least 5 likely old or young stars for each fractional abundance calculation, as defined in Section \ref{sec:popcf}. This produced a rough curve of $P_{mem}$ as a function of $D$, however it is not smooth due to small sample sizes in many bins, and also does not increase consistently with $D$, a result that would imply an unphysical pattern of $P_{mem}$ decreasing for stars closer to the cluster center. 

\begin{figure}
\centering
\includegraphics[width=8.5cm]{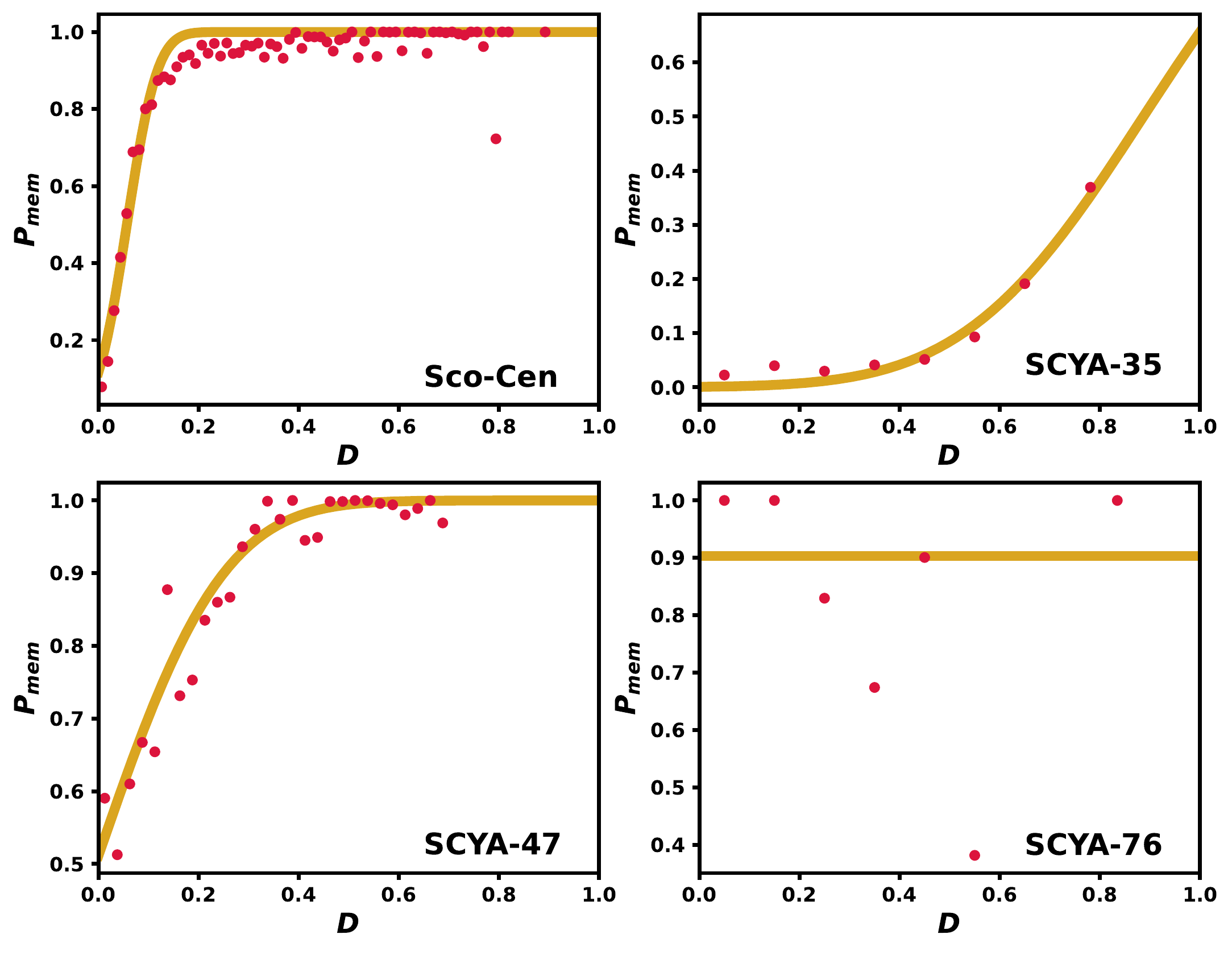}\hfill
\caption{Estimated $P_{mem}$ values for a range $D$, plotted as dots, alongside the fit we use as a map between $D$ and $P_{mem}$. The top row shows typical fits, with the top-left panel showing the young star-dominated Sco-Cen, and the top-left panel showing the field-dominated SCYA-35. The bottom row shows examples which include rare complications. For SCYA-47 in the bottom-left, its low stellar density as a function of $D$ over a wide range in $D$ resulted some outliers that skew the fit over $D=0.7$. The result shown restricts the fit to $D=0.7$, providing a strong agreement between the points and curve. The bottom-right panel shows SCYA-76, which is a group with so little field contribution that it does not produce a sensible curve. The fit provided is therefore just the average $P_{mem}$.}
\label{fig:dpmem}
\end{figure}

To smooth the result, we fit it with a Gaussian CDF. The wide range in cluster density profiles makes any profile fitting choices difficult to make universally, however Gaussian CDFs have asymptotes at 0 and 1, which should be the behavior of a membership probability curve. We also find close agreement between the Gaussian CDF profile and our most populated $P_{mem}$ curves. Some sample fits are shown in Figure \ref{fig:dpmem}. 

Using this fitting method, only 6 of 116 groups require additional attention to produce reasonable results. Four of these are group-dominated populations without enough field stars to produce a meaningful background contribution. For these, we simply calculated a mean $P_{mem}$ and attributed it to all members, and all but one of these values is between 85 and 91\%. The other is just below 60\%, and this is caused mainly by its age, which only marginally passes our 80 Myr isochronal age cut. This results in some more massive young stars that have already reached the main sequence being flagged as likely older than 50 Myr by our youth probability estimate. The remaining two groups had significant ranges with barely enough stars to permit $P_{mem}$ calculations, which resulted in difficult-to-fit noise spikes. In these cases, we restricted fitting to $D < 0.7$, and both produced very tight fits. The occasional assignment of stars in the older young associations to the likely old sample (see Sec \ref{sec:ages}) suggests that the $P_{mem}$ results of groups older than $~50$ Myr should be treated with caution, such as for Perseus OB3,  which has a gradual fit to its $D$--P$_{mem}$ conversion despite the visual dominance of the $\alpha$ Persei Cluster at high $D$.

These curves produce maps between $D$ and $P_{mem}$ for each individual group. For each star in a given group with a given $D$, we interpolated a value of $P_{mem}$ off the $P_{mem}$ vs $D$ curve for that group. Since $D$ is available for all stars in the minimally restricted stellar population, we were able to provide $P_{mem}$ values for all candidates in that extended sample. To ensure that all candidates have a credible chance of membership within its parent association, we used $P_{mem}$ to restrict the sample for each association, requiring that $P_{mem} > 0.05$. Not all groups contain stars that fail this $P_{mem}$ cut, so this change is only impactful in groups which have an especially weak separation from the field. This restriction is nonetheless important in those limited cases to ensure that the extents of groups do not incorporate too much of the field, which contains a background level of ejected association members and field binaries which become increasingly abundant for low values of $P_{mem}$ \citep[e.g.,][]{Sullivan21,Kerr22a}. Photometrically young stars used to define the group are kept regardless of whether they fail the $P_{mem}$ cut.

\begin{figure*}
  \centering 
  \includegraphics[width=17cm]{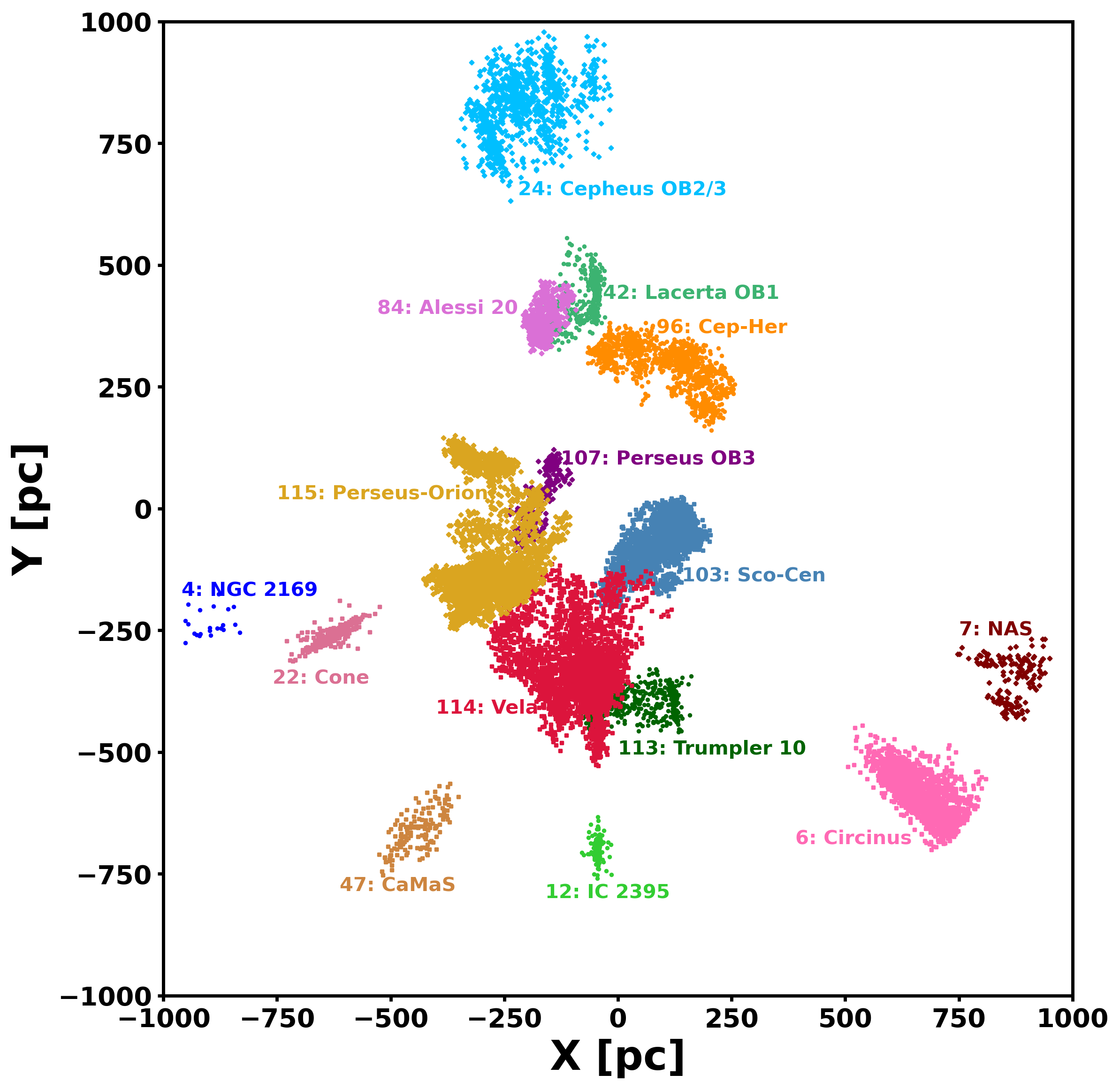}\hfill 
  \caption{Distribution of groups we identify in galactic coordinates, shown here in XY space, with all groups labelled. Stars shown are limited to young founding members. This panel shows the 15 groups with 8 or more O and B stars included in the founding population, which we use to indicate the most heavily populated groups.}
\end{figure*}
\begin{figure*}
  \ContinuedFloat 
  \centering \includegraphics[width=17cm]{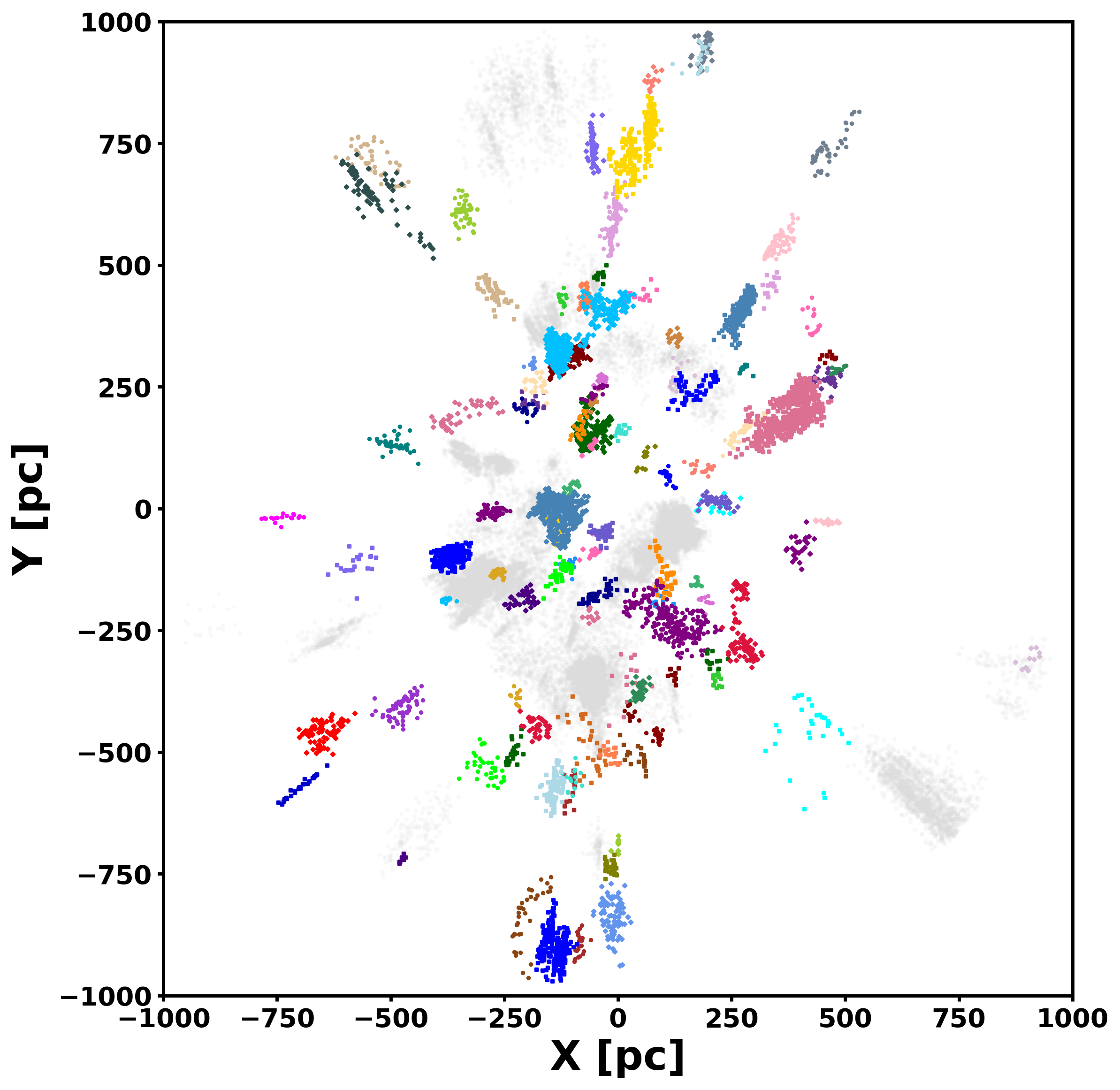}\hfill 
  \caption{(continued) The remaining 101 groups we identify, shown in XY galactic space. The labelled groups in the first panel are shown in the background in grey so that locations relative to these larger groups can be assessed without the larger groups obscuring the smaller ones. An interactive version in XYZ 3D galactic coordinates is available in the online-only version of this figure, which includes age data (see Section \ref{sec:ages}), as well as buttons to restrict the sample to interesting subsets of groups.}
  \label{fig:XYassocsmap}
\end{figure*} 

The resulting selection of candidate members includes 1.2 million stars across the 116 approved groups. The majority of these stars are expected to not be young, with the values of $P_{mem}$ providing an approximate likelihood of membership for each of these candidate members. In most cases, the total population of members identifiable with Gaia can be approximated as equal to the sum of all $P_{mem}$ values for stars in the association, which would infer a total population of real members in that set of approximately $2.8\times10^5$. A constant star formation rate over the 11.2 Gyr history \citep{Binney00} of the solar neighborhood would imply $2.5\times10^5$ stars formed in the last 30 Myr out of our total sample of 94 million, so the size of this population is consistent with expectations. 

The relationship between the number of identified stars and the corresponding population is set by the age and distance, which control recovery rates. While the sizes of the extended populations are in proportion to the size of the founding population in most cases, there are exceptions where inferred population sizes from $P_{mem}$ may differ significantly from the sizes of the populations accessible to Gaia. The extended population can be disproportionately large in situations where our detection of young stars happens in close proximity to older populations, such as in the case of SCYA-95, which produces extended populations far larger than the $P_{mem}$ curve should allow. The opposite effect is seen in older populations, where real members enter the set of likely old stars used to set $P_{mem}$. This can result in a systematic underestimation for $P_{mem}$. While these effects warrant caution, particularly when using our lists for statistical assertions, effects skewing $P_{mem}$ appear to be quite rare in our sample, especially in nearer and more substantial populations. 

\input{Memberlist.tex}

We list credible candidate members of each association in Table \ref{tab:allpops}, consisting of both the population of young stars used to define the group in the clustering step, and the extended populations with $P_{mem} > 0.05$. We include stars that do not pass the astrometric and photometric quality restrictions, and include relevant flags necessary for re-imposing these restrictions, including the astrometric and photometric quality flags, $\pi/\sigma_{\pi}$, $P_{mem}$, and whether the star was in the young set used to define the group. We also provide basic information for each prospective member, including the Gaia ID, position, and magnitude. In Table \ref{tab:popsum} we provide a summary of all notable stellar samples referenced throughout this paper, which serves as an overview of the operations that led to the final sample shown in Table \ref{tab:allpops}.

\begin{figure*}
  \centering 
  \includegraphics[width=17cm]{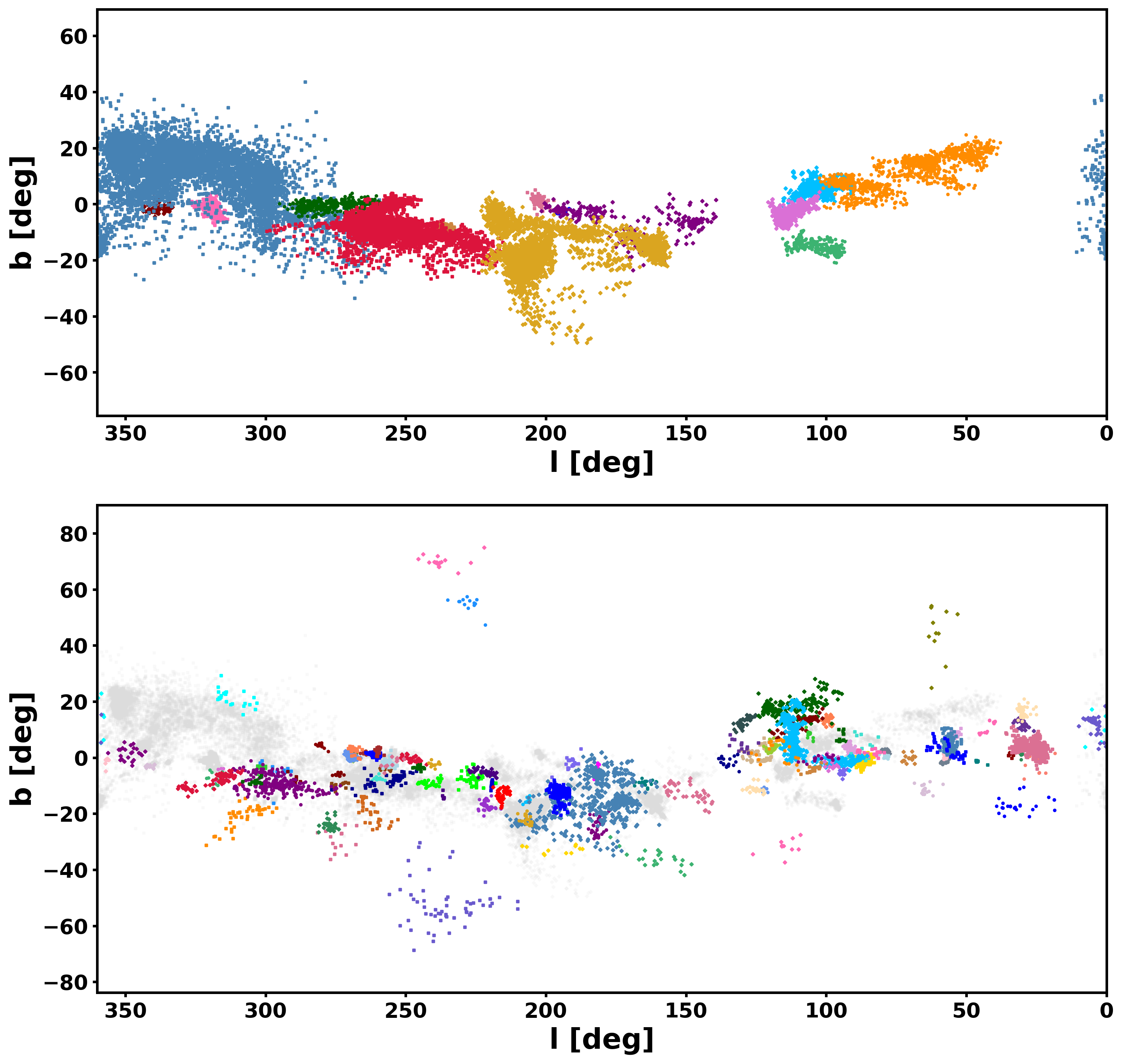}\hfill 
  \caption{The distribution of groups in l/b galactic sky coordinates. Our plotting choices are the same as for Figure \ref{fig:XYassocsmap}, with the top panel showing the large groups with 8 or more OB stars, and the bottom panel showing all other groups, with the larger groups in grey. An interactive form of this plot is available in the online-only version of this figure, which paints groups by their distance.}
  \label{fig:lbassocsmap}
\end{figure*}

\begin{figure*}
  \centering 
  \includegraphics[width=17cm]{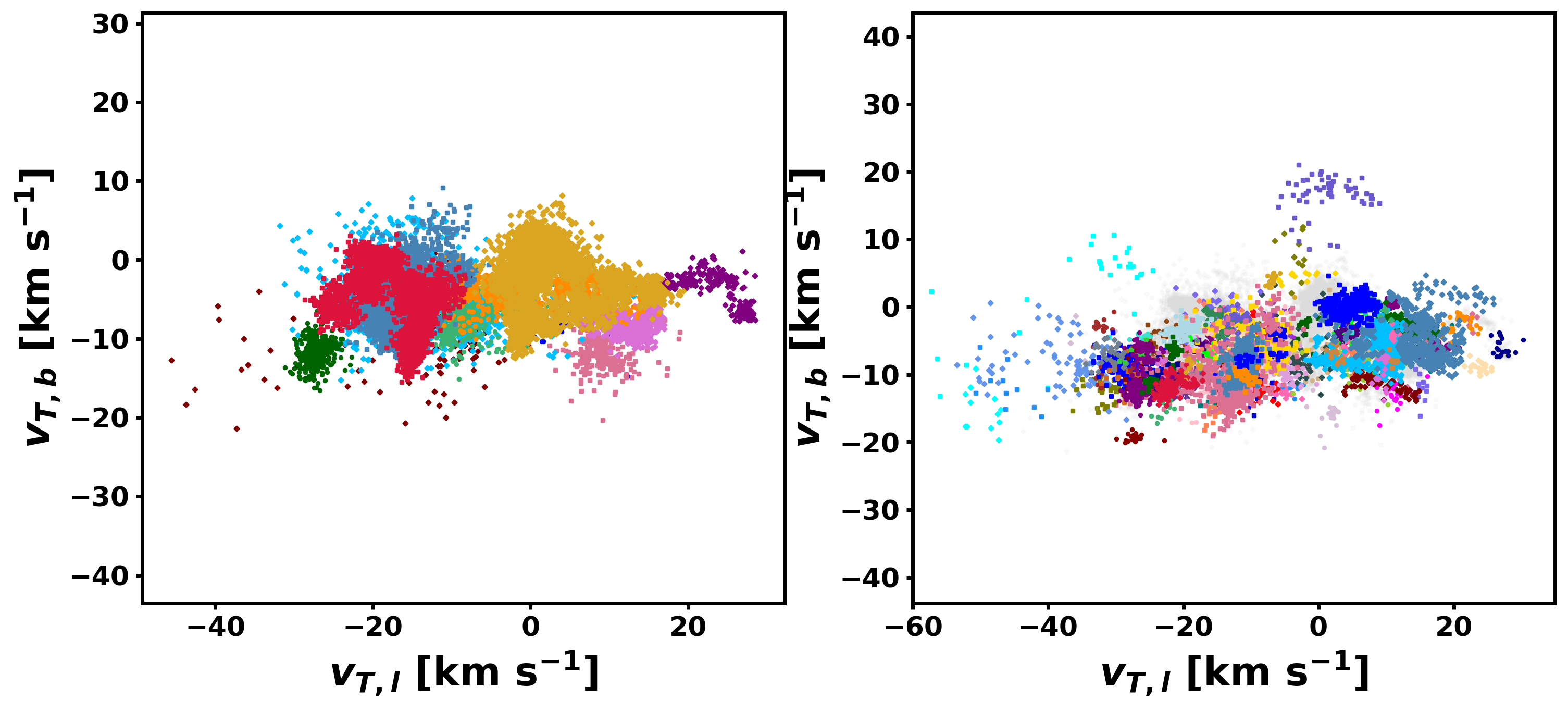}\hfill 
  \caption{The same as Figures \ref{fig:XYassocsmap} and \ref{fig:lbassocsmap}, but in transverse velocity. The left panel shows groups with 8 or more OB stars, and the right panel shows all other groups. An interactive version of this plot is available in the online-only version of this figure. There we also have selector buttons that limit the sample to either newly-discovered groups with higher average velocities (see Section \ref{sec:microgroups}), or the Circinus complex, which is shown separately in Figure \ref{fig:vtcircinus}.}
  \label{fig:vtassocsmap}
\end{figure*}

The distributions of groups in Galactic Cartesian coordinates, Galactic sky coordinates, and transverse velocities are shown in Figures \ref{fig:XYassocsmap}, \ref{fig:lbassocsmap}, and \ref{fig:vtassocsmap}, respectively. To improve the view and highlight the most notable associations, we plot groups with 8 or more OB stars in the founding population separately, and provide a second panel for groups with less than 8 O an B stars. This includes many well-known groups as well as a few with limited coverage in the literature, which are discussed in Section \ref{sec:groupid}. The Circinus complex is a peculiar case in transverse velocity, so we discuss it separately in Appendix \ref{app:circinus}. We provide interactive versions of all three plots in the online-only version of this paper, allowing for subsets of the data to be selected as desired, alongside age measurements, which are introduced in Section \ref{sec:ages}.

\begin{table*}
\centering
\caption{Summary of populations discussed in this paper, including their sizes, an explanation of their origin, and the section of this paper they originate from. The sample in Table \ref{tab:allpops} is given by the ``Final Cluster Candidate Sample''. }
\label{tab:popsum}
\begin{tabular}{cccc}
\toprule
Sample & Number of Stars & Description & Section\\
\midrule
Unrestricted (UR) & 94,238,210 & Full Gaia sample & \ref{sec:data}\\
Quality-Restricted (QR) &  52,965,232 & UR with poor measurements removed & \ref{sec:data}\\
Photometrically Young (PY) &  418,611 & QR with P$_{Age<50 Myr} > 0.2$ & \ref{sec:ysselect}\\
High-Quality Young (HQY) &  181,524 & PY with restrictions to improve clustering & \ref{sec:clustering}\\
Young Clustered (YC) & 38,899 & HQY stars located in a cluster & \ref{sec:clustering}\\
Extended Clustered (XC) & 3,053,874 & YC + phase-space neighbors from UR  & \ref{sec:extpops}\\
Vetted Young Clustered (VYC) & 36,182 & YC with false groups vetted  & \ref{sec:vetting}\\ 
Extended Vetted Clusters (XVC) & 1,317,609 & XC with false groups vetted  & \ref{sec:vetting}\\ 
Final Cluster Candidate Sample  & 1,222,152 & XVC with low-$P_{mem}$ stars outside VYC removed & \ref{sec:dpmem}\\ 
\bottomrule
\end{tabular}
\end{table*}

\section{Survey and Cluster Overview} \label{sec:results}

With 116 associations detected, our updated census provides a significant expansion over SPYGLASS-I, which found only 27 groups at the top level, or 26 excluding the older Pleiades cluster. Much of this expansion comes from the widening of our survey to a 1 kpc search horizon, although we also see significant deepening of our survey within the 333 pc search horizon of SPYGLASS-I. Of the 116 groups in our sample, 41 have at least 10 stars within 333 pc, making them detectable using only the stars within that radius. We therefore nearly double the number of groups available relative to SPYGLASS-I, even when that original search radius is considered. The increase in survey depth is even more notable when we consider group mergers. The higher sensitivity in this work often results in low-density populations connecting adjoining associations (see Section \ref{sec:lsm}), such that the 26 young populations in SPYGLASS-I merge to form only 16 populations here. As a result, this survey expands to the number of populations represented within 333 pc by a factor of 2.6, and expands the total list by more than a factor of seven.

Since much of the 30 Myr old sequence has already descended close to the main sequence, nearly all of the identifiably young stars are M dwarfs. As such, unless the association is large enough to be identified purely based on O and B stars, our methodology loses the ability to detect new structures with age $\tau \gtrsim 25$ Myr at about 400-500 pc, where Gaia begins lose sensitivity and astrometric quality at the bottom of the pre-main sequence. Therefore, despite the rich set of stellar populations identified here, there is likely significant structure in the outer reaches of our search radius waiting to be discovered. 

In this section we provide basic information on the populations identified in this work. This includes both a broad overview of the position the populations we identify hold within the literature, and their intrinsic properties, allowing each association to be contextualized within our current knowledge of associations.  

\subsection{Our Groups in Literature} \label{sec:groupid}

Most of the associations we identify in this work are known in the literature. We therefore must cross-match our lists with known populations to locate existing identifiers that better contextualize these groups. For better-established groups we simply provide common names, while some of the smaller groups required direct cross-matching with additional lists. We compared our catalogs to the Theia and UPK lists \citep{Kounkel19, Sim19}, as well as the catalogs from \citet{CantatGaudin20} and \citet[][]{Prisinzano22}, which represent four of the deepest existing surveys covering young stellar populations. We initially flagged groups with any stars cross-listed with any of these catalog associations, and then investigated each possible match individually to determine whether the match has both clear overlap with the core of our group, as well as a similar extent. Those with reasonable matches have their catalog IDs provided as their name in Table \ref{tab:groupstats}. Of the 116 groups identified, 74 had clear equivalents, with the remaining groups either overlapping with but clearly different from a known group, or completely unknown. 

Cases of overlapping but non-equivalent populations were most common in comparisons with the \citet{Kounkel19} catalog, as our populations were often grouped together under large-scale ``string'' structures rather than separate populations. We have our own independent merging condition through the use of $\epsilon$ in our HDBSCAN implementation, which merges structures by identifying young stellar populations in between that bridge disparate overdensities into continuous structures. While many groups remained independent from each other in our clustering results, this merging condition did cause some mergers at scales comparable to or exceeding those of \citet{Kounkel19}, including the merging of the widely-separated populations of Perseus OB2 and the Orion Nebula Complex (see Section \ref{sec:lsm}). Despite frequent group mergers in our sample, some of the ``strings'' in the Theia catalog contain many SCYA groups. While our sensitivity to late K and M stars drops at $\sim 400-500$ pc, resulting in the potential loss of some connecting structure, many of the Theia strings containing the most SCYA groups are within that radius, suggesting that these groups' lack of a merger is not a sensitivity issue. As a result, we see little reason for further agglomeration of groups, and therefore treat independent populations within Theia groups as new populations, rather than components of a known agglomeration. However, we do note groups with overlap in Table \ref{tab:groupstats}. This choice is supported by follow-up studies which have shown a lack of consistent velocity coherence in these strings, further questioning whether the sub-components of these structures have genuine connections to one another \citep{Zucker22a}.

Out of the 116 groups we find, we categorize 27 as overlapping with known populations, but having a different enough extent that they cannot be directly connected to any known population. One of these groups, SCYA-47, is substantial enough that it is included in the set of large groups with 8 or more OB stars shown in Figures \ref{fig:XYassocsmap}, \ref{fig:lbassocsmap}, and \ref{fig:vtassocsmap}. It includes components of Theia 86 and 87 from the \citet{Kounkel19} catalog as well as parts of \citet{Prisinzano22} group 633, although none of these groups resemble the extent we show. Due to its status as a new and substantial population, we refer to this group as the Canis Major South (CaMaS) Association for future reference. SCYA-7, which contains the open clusters NGC 6250 and NGC 6178, was not counted as unknown in its presented extent since it is dominated by known structures, however the clusters it contains have not been previously connected. Due to the scale of the structure, we name it the Norma-Ara-Scorpius (NAS) Complex for future reference, after its location near the tripoint between those constellations. In addition to the 27 groups with weak connections to known populations, 15 groups had no recognizable equivalents across these catalogs. 

While this paper was under review, a new catalog which uses HDBSCAN clustering on an age-unrestricted DR3 dataset was published by \citet{Hunt23}, providing one of the most sensitive general cluster surveys to date. This paper independently discovered 5 of the 15 previously-undiscovered groups. Of the groups with weak connections to known populations, 7 of 27 have close matches with the \citet{Hunt23} catalog, with most of the remaining populations having at least some overlap. However, despite this survey's sensitivity, we found that 14 of our populations had no equivalent in the \citet{Hunt23} catalog, demonstrating that an age-restricted survey may be necessary to detect the most tenuous young groups. 

After the inclusion of the DR3-based \citet{Hunt23} cluster survey, 10 of our groups are entirely unknown, and an additional 20 have no direct equivalent in the literature. This publication therefore provides a significant deepening of our knowledge of young associations over not just SPYGLASS-I, but also the existing literature. The properties of our newly-discovered groups in spatial and velocity coordinates are often unique, and we discuss the properties of these groups in further detail in Section \ref{sec:microgroups}. 

\subsection{Ages} \label{sec:ages}

While stellar ages can be calculated directly from the SPYGLASS infrastructure (as outlined in SPYGLASS-I), ages for individual stars can be impacted drastically by local inaccuracies in the reddening corrections, along with typical scatter in magnitude and distance. Computing ages on an association level can smooth out many of these systematic variations, resulting in a more reliable solution. While many groups contain significant internal age variations such as those shown in SPYGLASS-II, association-level ages nonetheless provide a broad age categorization for the group, which can be later broken into sub-populations for studies on the association level. We therefore only provide association-level ages in this publication.

Our age calculation roughly following the method employed in SPYGLASS-I, using a restricted form of the extended population for isochrone fitting. The quality restrictions we employ are selected to minimize contamination from non-members and binaries. First we required that all stars in our fits have $P_{mem}>0.9$ and RUWE$<$1.1, removing most field contamination and binaries, respectively. This cut on RUWE is quite harsh, and was used in SPYGLASS-II to produce a sample nearly free of unresolved binaries for fitting, at the expense of completeness \citep{Bryson20}. We then restricted our sample to the pre-main sequence by requiring that $1.2<G_{BP}-G_{RP}<4$ and $G>3$, which provides a region where age varies with magnitude in a predictable and consistent manner. An additional vertical cut to limit field contamination was also included, requiring that all stars be above an 80 Myr solar-metallicity PARSEC isochrone. This requirement was omitted for any older populations with part of their sequences below this isochrone. The lower sensitivity of our star detection for older populations means that groups in this category tend to dominate their parameter space, typically making broader selections appropriate for capturing the group without a considerable increase in contamination.

\begin{figure}
\centering
\includegraphics[width=7.5cm]{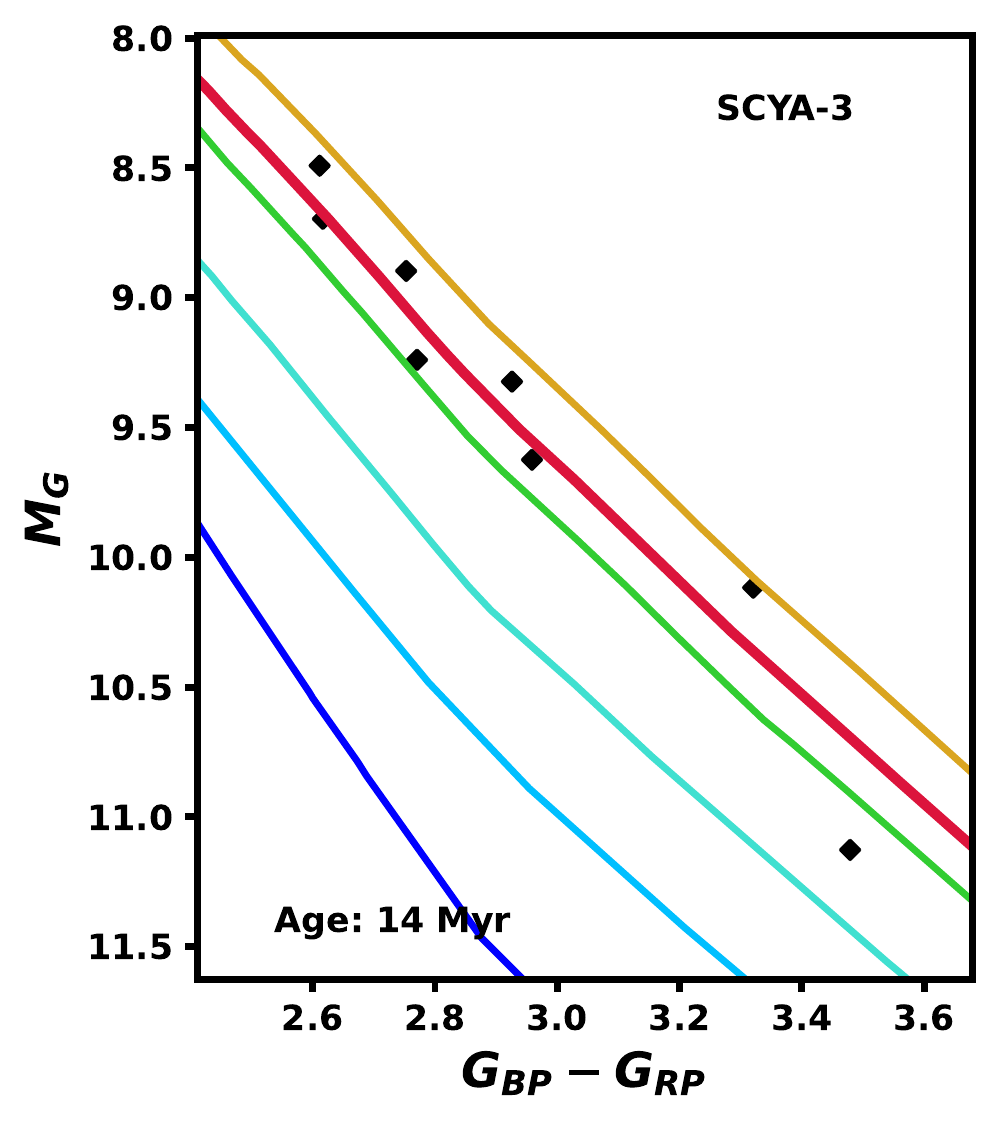}\hfill
\caption{A sample age fit, showing SCYA-3 (H-R 3321). The stars included in the fit are shown as black diamonds, and the best fit pre-main sequence isochrone is shown as a thick red line. The stars included follow the restrictions described in Section \ref{sec:ages}, which limit the contamination from binaries and field stars. Isochrones of 10 Myr, 20 Myr, 40 Myr, 80 Myr, and 1 Gyr, from top to bottom, are shown for reference. The full set of fits for all 116 SCYA associations is provided in the online version of this paper.}
\label{fig:ageset}
\end{figure}

The restrictions presented so far represent a harsh set of limits, especially the requirement that $P_{mem}>0.9$, a condition which members of some associations never pass. We therefore provide loosened restrictions that allow age calculation for smaller and more tenuous groups. For groups with less than 8 stars that meet our initial requirements, we first relax the binarity cut to RUWE$< 1.2$, a cut used in SPYGLASS-I and SPYGLASS-III in cases where completeness is desired over purity. For any sample that still has less than 8 members under that loosened restriction, we simply select the 8 members with the highest $P_{mem}$ out of a sample that satisfies our color and magnitude cuts and has RUWE$< 1.2$. While samples with less restrictive cuts on $P_{mem}$ are much more prone to contamination, we find that our cut on the 80 Myr isochrone does enough to limit field contamination that reasonable age solutions are still attainable. Once the samples for each group were selected, we used a least-squares fitting algorithm to fit the stars in color-magnitude space against a grid of PARSEC isochrones from 1 to 80 Myr. 

The resulting age fits are shown in Figure \ref{fig:ageset}, showing the high-confidence stars used for fitting alongside the best-fit isochrone. We do not include any uncertainties in our results due to the likely presence of substructure within these populations. The most obvious of these is the Vela complex, which shows a very clear double sequence in its CMD, however there are many other examples such as Sco-Cen and Perseus-Orion which also show evidence for non-coeval substructure. However, due to the scale of this survey, the complete sub-clustering and age-dating of all substructures is beyond the scope of this publication. More substantial populations with extensive substructure will therefore require regional studies to provide a comprehensive view of their history. However, for smaller associations without significant internal structure, these ages should be much more widely applicable. 

\subsection{Other Bulk Group Properties}

Aside from age, we compile numerous other cluster properties in Table \ref{tab:groupstats}, following the properties included in the tables in SPYGLASS-I. These include median sky positions in galactic and celestial coordinates, as well as median distance, proper motion, and transverse velocity. For distance we also include a standard deviation, which can be indicative of the radial extent of associations at short distances, but tends to be increasingly dominated by parallax uncertainty towards the edge of our search radius. We also include approximate on-sky extents and velocity extents. The values reported are major and minor axes fit from a multivariate Gaussians, and come from the galactic l/b coordinates and $v_{T,l}$/$v_{T,b}$ transverse velocity coordinates, respectively. 

\input{AllGroups.tex}

\section{Notable Features} \label{sec:discussion}

The results presented in Section \ref{sec:results} cover a range of associations with widely varied positions, velocities, and sizes. A substantial share of these populations are either little-studied or completely unknown, so these populations provide important opportunities for future research and discoveries related to young stars, their formation mechanisms, and the properties of the planetary systems they contain. In this section, we highlight some of the most notable features to emerge from our analysis, and discuss their potential implications. The results we discuss are drawn directly from the data summaries provided in Table \ref{tab:allpops}, the group overviews in Table \ref{tab:groupstats}, and the visual representations provided in Figures \ref{fig:XYassocsmap}, \ref{fig:lbassocsmap}, and \ref{fig:vtassocsmap}. This provides a showcase of potential future research avenues which the study of our associations would facilitate. 

\subsection{Large-Scale Mergers} \label{sec:lsm}

Since our clustering was designed to identify regions of interest for studies of common star formation, we merged some regions which are not typically merged. The most notable examples of this are the Vela complex, which includes a few dynamically different subregions described in \citet{CantatGaudin19}, and the Orion-Perseus complex, which contains the Orion Molecular Cloud complex, Perseus OB2, and the recently-discovered but still notable Monoceros Southwest region described in SPYGLASS-I, in addition to weakly-defined connecting structure. Sco-Cen also grows in this survey, merging its extent from SPYGLASS-I with the main Chamaeleon Complex (i.e., Cha I and II). The last substantial new merger in this publication is SCYA-96, which contains the populations of Lyra, Cerberus, and Cepheus-Cygnus from SPYGLASS-I in addition to previously-known and more distant $\delta$ Lyrae and RSG-5 clusters \citep{Stephenson59, Roser16}. The combined population contains over 1000 stars spanning nearly 300 pc. We name this population Cepheus-Hercules, or Cep-Her\footnote{pronounced to rhyme with ``Zephyr''}, after the endpoints of the constellations it spans, and we discuss the potential implications of this population and other similar structures in Section \ref{sec:cepher}. 

While components of all other major merged populations have as least some history of being studied together (see SPYGLASS-I for Sco-Cen, \citealt{CantatGaudin19} for Vela, and \citealt{Kounkel19} for Cep-Her), the merged Orion--Perseus association (SCYA-115) is unique in the range of populations it includes, with the component Orion Nebula Complex, Perseus OB2, and even Monoceros Southwest complex all having been consistently viewed as separate to date, even by works like \citet{Kounkel19} which routinely merge smaller-scale structures. These populations are merged despite the exclusion of the $\lambda$ Orionis Cluster, which is occasionally discussed alongside the Orion complex \citep[e.g.,][]{Zari19}. The reason for their merger in this work is the presence of tenuous young stellar populations filling the space between the populations, largely consisting of stars identified as part of Taurus-Orion IV in SPYGLASS-I.

Despite the disparate positions of the core populations being merged, the components of this combined population actually have very similar ages to one another in SPYGLASS-I, with both Perseus and Orion having regions of active star formation alongside older generations with maximum ages around 20 Myr old. Monoceros Southwest is older at $\sim$ 25 Myr, but not far from the range seen for the other populations. This suggests that the entire complex could be explained by a single, initially co-spatial event in which star formation begins in a large-scale molecular cloud which breaks up, likely under the influence of feedback from O and B stars, eventually resulting in the emergence of more significant star formation events in Perseus, Orion, and Monoceros Southwest, while leftover material in between produces smaller star-forming events that link the structure. More thorough traceback and age-dating will however be necessary to assess these connecting structures and establish whether they are dynamically consistent with being bridge structures between these larger populations. This work will be necessary to confirm these connections, as the low velocity between the complex and the field makes Orion particularly vulnerable to field contamination, which may expand group boundaries and falsely merge structures. 

\subsection{Cep-Her and Evidence for Large-Scale Patterns} \label{sec:cepher}

\begin{figure*}
\centering
\includegraphics[width=19cm]{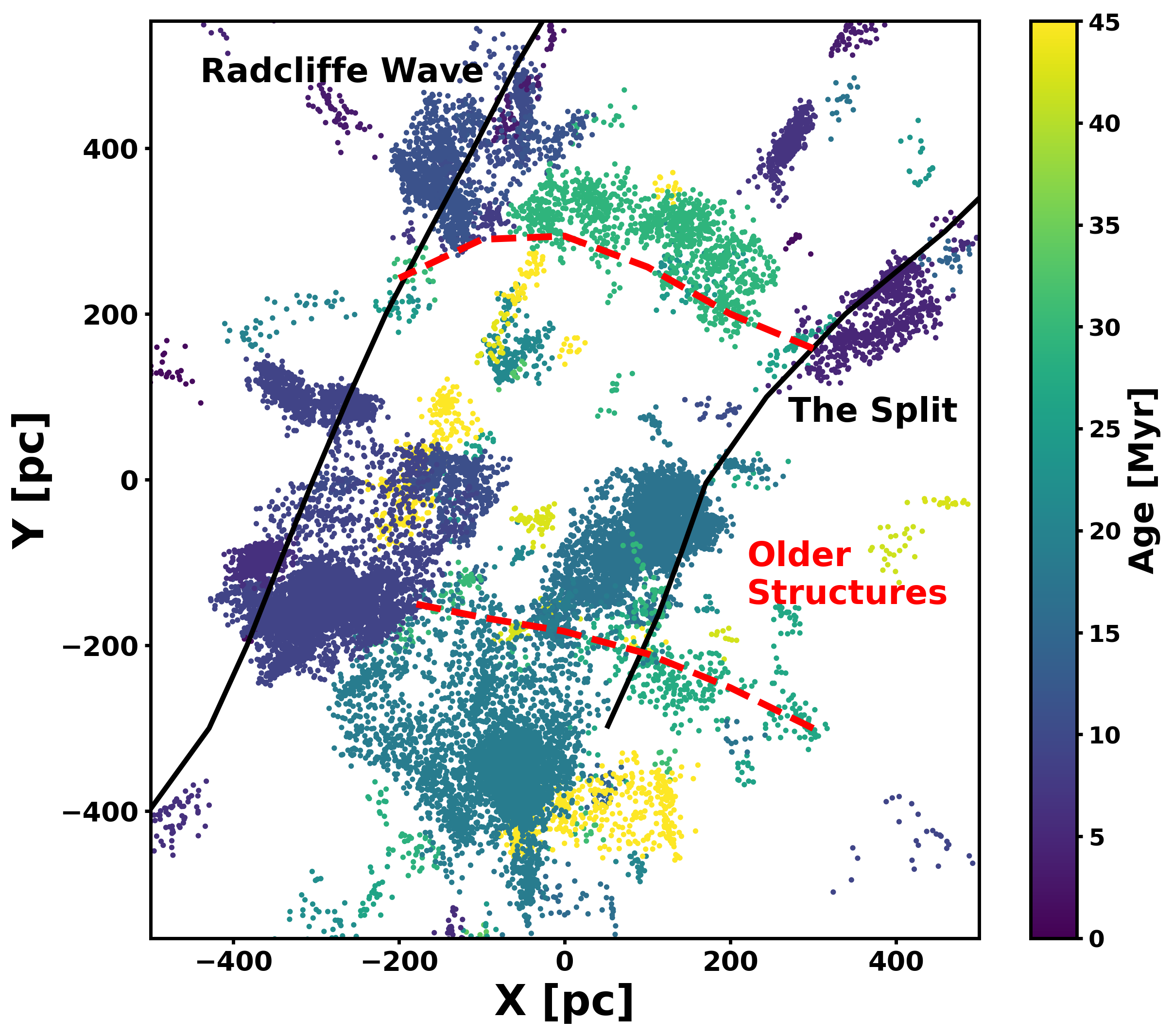}\hfill
\caption{The distribution of ages for stellar populations, presented in the XY plane, and limited to the inner 500 pc where patterns are visible for stars near 30 Myr. Annotations show the locations of the Radcliffe Wave, the Split, and our proposed older structures. The age distribution can also be interacted with in the online-only version of Figure \ref{fig:XYassocsmap}, which allows the isolation of structures around 30 Myr, or the young structures (age $<20$ Myr), which are plotted alongside the outlines of the Radcliffe Wave and the Split. }
\label{fig:agedist}
\end{figure*}

\begin{figure*}
\centering
\includegraphics[width=16cm]{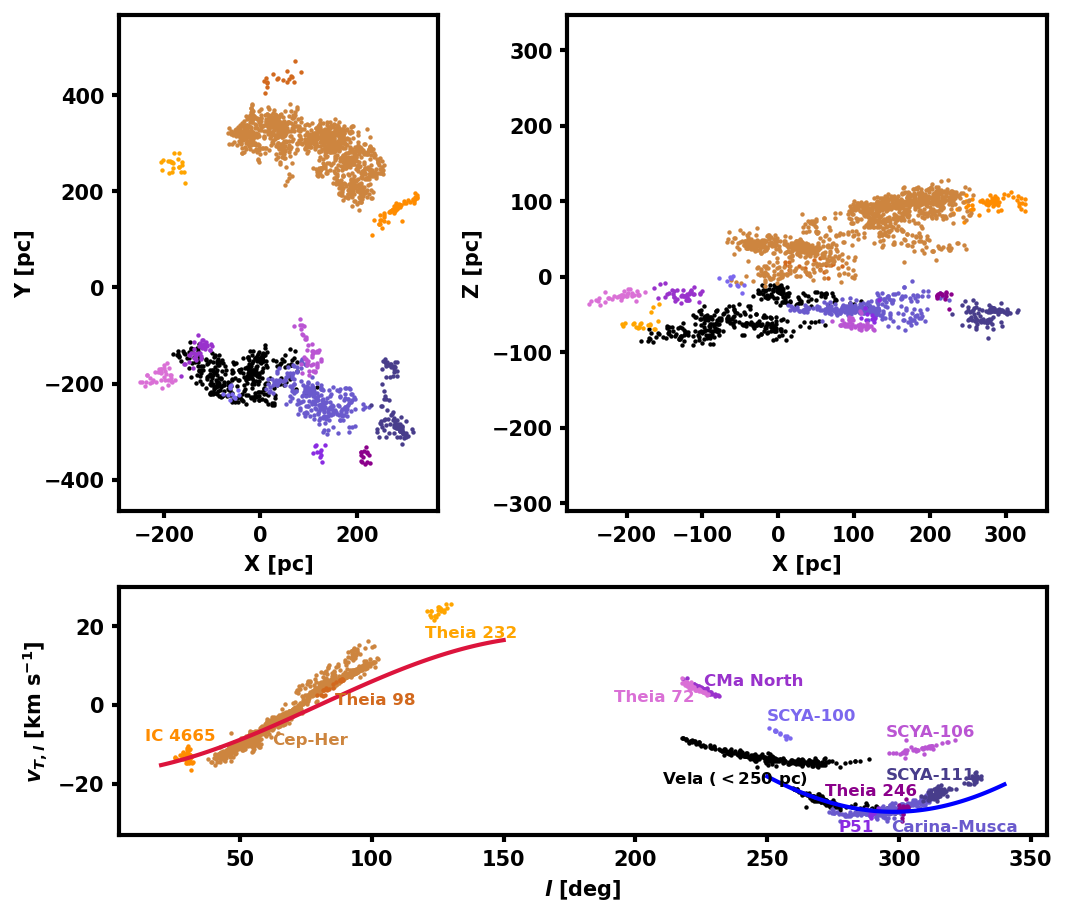}\hfill
\caption{Component populations of proposed 25-35 Myr structures, which are extended along an axis perpendicular to known spiral structure. We display these structures in XY and XZ galactic coordinates, showing their spatial coherence, as well as $l$-$v_{T,l}$ space, in which continuous sinusoidal arcs typically indicate common velocities which are modulated by projection effects in $l$. Shades of orange are associated with Cep-Her, while shades of purple and blue are associated with Carina-Musca. We also include the near edge of the Vela complex in black, which has a similar age and location to Carina-Musca. Both proposed structures follow consistent arcs in both XY and XZ coordinates, with Cep-Her's proposed structure being inclined to the galactic plane, while Carina-Musca's structure is relatively flat. Cep-Her and its companions show a tight velocity trend in $l$, consistent with projections of a single common velocity. Carina-Musca has multiple components in $l$-$v_{T,l}$ space, indicating that it is less dynamically coherent, but it still shows a few continuous arcs, with the largest consisting of Carina-Musca, SCYA-111, P51, Theia 246, and part of Vela, which is similar to Cep-Her in its spatial extent. We show the projected median 3D velocity vectors of the $\delta$ Lyrae cluster in Cep-Her and Carina-Musca as red and blue curves, respectively, demonstrating relatively consistent velocities of these groups. These structures are prime targets for RV followup, which would enable the reconstruction of patterns in their formation sites via traceback.}
\label{fig:spurs}
\end{figure*}

The Cep-Her complex (SCYA-96) is the fifth-most populated association we identify in this survey, and it is also one of the largest, with an end-to-end spatial extent comparable to Sco-Cen. This is not the first time a large population in this region has been proposed. The structure was first broadly identified on stellar density maps by \citet{Zari18} and later presented by \citet{Kounkel19} as the 3000-member Theia 73, which contains many of the populations included by HDBSCAN in this work. However, Theia 73 is not identical to our definition of Cep-Her, as it skips many of the sub-populations closer to the galactic equator, while including populations like CFN which our HDBSCAN implementation sees as completely separate. The region containing Cep-Her is also noted by \citet{Prisinzano22}, where it is labelled as connected to the $\delta$ Lyr cluster. 

In this work, Cep-Her contains 1164 founding young members. Given the lower sensitivity of our young star identification for ages near Cep-Her's bulk age of 29 Myr, we expect a low recovery rate for members in the range of 20-25\% (see the completeness discussion in SPYGLASS-I), implying a total population of around 5000. This puts Cep-Her on a similar scale to many of the other great young associations in the solar neighborhood, and its age makes it a potential older analog to populations like Sco-Cen. Cep-Her also contains some of the youngest stars in the Kepler field, and planets have already been found around multiple stars within, making Cep-Her of particular interest for studies of young planets \citep{Bouma22}. However, the large scale of Cep-Her may make it particularly important as a probe of the large-scale star-forming structures present at the time of its formation.

The study of large-scale star-forming structures has advanced considerably in recent years with the discovery of the Radcliffe Wave \citep{Alves20} and the Split \citep{Lallement19}, which were identified as kpc-scale dust overdensities. Both of these structures include stellar populations alongside this gas and dust, which comprise most major young associations in our sample \citep{Zucker22}. We show the spatial distribution of stars in our survey within the nearest 500 pc in Figure \ref{fig:agedist}, coloring groups by the ages computed in Section \ref{sec:ages}. Components of the Radcliffe Wave are traced by a chain of substantial associations, which all have ages younger than 10 Myr. Our view of the Split contains a wider range in ages due to the older populations in Sco-Cen and Vela which exist alongside regions of active formation, but Sco-Cen, Vela, and Serpens still provide a strong outline for the local extent of this structure. The young ages of the populations associated with the Split and the Radcliffe Wave reflect their discovery through dust overdensities, which disperse soon after star formation. O and B stars have been used to identify similar structures without the presence of gas, such as the Cepheus Spur \citep{PantaleoniGonzalez21}, however the potential for discoveries using this method are limited by the rarity of these luminous stars. The discovery of the Cep-Her therefore provides an older complement to the younger structures which comprise the Radcliffe Wave and the Split. The newfound accessibility of these older populations allows us to begin tracing the evolution of star formation not just within populations, but as a continuous system of interconnected processes that facilitate the initiation, progression, and termination of star formation throughout the nearby spiral arms.

Cep-Her's age provides ample time for the warping and dispersal of its population, so conclusions based on the current configuration of the association should be treated with caution, pending dynamical traceback studies.  Nonetheless, the current forms of Cep-Her and some other nearby groups show patterns which are unique compared to the younger structures, most notably their spatial orientations. Unlike most large nearby structures, which either follow the nearby spiral arms or diverge from them at a very acute angle \citep{Zucker22}, Cep-Her has a near-perpendicular orientation, spanning the 400 pc gap between the Radcliffe Wave and the Split while having a width of only about 100 pc. Similarly-aged but separate populations are also found on either side of Cep-Her, accentuating this pattern and bringing the total length of this possible structure to over 500 pc. 

We show the spatial distribution of the potential component populations in Figure \ref{fig:spurs}, alongside the variation in $v_{T,l}$ transverse velocity with galactic $l$ coordinate, which we use to indicate whether velocity spreads are consistent with geometric projection effects. There we show that the proposed component populations lie within a common plane in 3D space which is inclined to the galactic plane. Most velocity variations within these populations occur along the $v_{T,l}$ axis, and this variation has a very consistent trend in $l$, indicating that geometric projection effects dominate the velocity differences that we see. To verify this, we calculated projected radial velocities across Cep-Her using the known radial velocities and proper motions of the $\delta$ Lyrae cluster, a centrally-located component of Cep-Her \citep{CantatGaudin20b, Tarricq21}. The result shows that mean velocities in Cep-Her, IC 4665, and Theia 98 all stay within 5 km/s of this velocity vector throughout their extents in $l$, limiting the potential for structural change in the time since its formation.

This perpendicular orientation for associations around 30 Myr is not unique to Cep-Her and its neighbors, as another chain of $\sim$30 Myr old groups can be found opposite the Sun from that complex, stretching from Canis Major North (SCYA-65) to SCYA-111 and containing Carina-Musca. The subregions of Vela and Sco-Cen that overlap with this chain of associations, namely the CG4 subgroup in Vela \citep{CantatGaudin19} and the IC 2602 branch in Sco-Cen, are only slightly older than these other structures, with SPYGLASS-I ages between 30 and 45 Myr, indicating a possible extension to this age pattern. Like in Cep-Her, all potential component groups lie along the same plane in 3-D galactic coordinates, however they are less clearly connected in velocity coordinates, with a few distinct arcs emerging in $l$-$v_{T,l}$ space (see Figure \ref{fig:spurs}). While this does not conclusively rule out a connection between these groups, it does emphasize the importance of a full 3-D dynamical analysis, which can establish whether these dynamically-distinct components trace back to any coherent past structure. However, even if individual arcs in $l$-$v_{T,l}$ space are assumed to be separate, we still see a 400 pc-long structure in the region with velocity coherence, consisting primarily of Carina-Musca, SCYA-111, and part of Vela. We show a projected velocity vector for Carina-Musca in Figure \ref{fig:spurs}, which is calculated using Gaia RVs, proper motions, and positions for stars in our sample with $P_{mem}>0.8$, and like in Cep-Her, the differences between this velocity vector and mean transverse velocities in these populations remain less than 5 km/s throughout its extent. Regardless of its exact parameters, the presence of another structure in addition to Cep-Her may nonetheless indicate a pattern of star formation along a perpendicular mode which dominated local star formation between 25 and 35 Myr ago. This motivates follow-up dynamical studies to confirm that these groups were both coherent and maintained orientations perpendicular to the spiral arms at the time of formation. 

If the current pattern of arm-perpendicular structures holds when traced back to formation, it would indicate that star formation in the solar neighborhood recently followed a pattern dominated by spurs, which are defined as chains of OB associations that deviate from the main arm at large angles \citep{LaVigne06}. These structures are often seen cutting between spiral arms in other galaxies, and extend outwards from the primary dust lanes and young stars that trace the spiral arms \citep[e.g.,][]{LaVigne06}. Spurs are also often found to have relatively regular disk density-dependent spacing of between 100-700 pc in both observations and models, which is quite comparable to the 400-500 pc between the structures we propose here \citep[e.g.,][]{Wada04, Dobbs06, LaVigne06}. These structures therefore appear consistent with established patterns in galaxies, and could represent a strong local equivalent. However, we also know that the patterns of local star formation in the last 10-20 Myr appear to be dominated by large-scale structure with pitch angles consistent with spiral arms \citep{Zucker22}. This suggests that the sun has recently seen a transition in nearby large-scale star formation patterns, moving from a mode dominated by structures perpendicular to the arm to a mode dominated by kpc-scale arm-aligned structures within the last 30 Myr. This further motivates more large-scale mapping, traceback, and age dating within the solar neighborhood to better understand this change, and established whether it is due to the evolution of local structure, or just the movement of gas-decoupled associations through different sections of the spiral arm. Young associations have the unique capability to trace star formation in not just space, but also time, enabling a view of the evolution of these structures which could provide critical comparisons to models of spiral arm development and evolution \citep[e.g.,][]{Shetty06, Kim20}.

\subsection{Small, High-Velocity, and High-Latitude Groups} \label{sec:microgroups}

Of the populations we detect here, 15 of them were unknown prior to Gaia DR3 \citep[e.g.,][]{Kounkel19, Sim19, Prisinzano22}. All of these groups have internal velocity scatters consistent with other groups in our sample (see Table \ref{tab:groupstats}, Figure \ref{fig:vtassocsmap}), which are not consistent with the typical scatter of field populations (see Appendix \ref{app:circinus}). Some have pre-main sequences that clearly separate from the field, while others are much more tenuous, making spectroscopic follow-up necessary to confirm their existence (see the full figure set for Figure \ref{fig:clustervetting}).  The most tenuous populations can have fewer than 10 pre-main sequence stars that visibly separate from the main sequence for some range of $P_{mem}$, with values of $P_{mem}$ that never exceed 0.3.  These groups are therefore near the limit of detectability using a purely photometric and astrometric survey. The inclusion of additional radial velocities on a mass scale would help to limit the field contamination during clustering, allowing more tenuous structures to be detected. However, with Gaia RVs generally limited to magnitudes $G<14$ and often containing large uncertainties \citep{GaiaDR322}, such an update would require generational improvements to our RV measurement infrastructure. Nonetheless, if spectroscopically confirmed, these groups would represent a substantial new reservoir of young associations, significantly expanding our knowledge of recent star formation. 

Many of these populations are noteworthy for their unique properties relative to some of the better-known large associations. The mean transverse velocities of these  groups in particular are often quite anomalous. SCYA-26 and SCYA-28 both have combined transverse velocities in excess of 50 km s$^{-1}$, and combined nearly half of these small new associations have transverse velocities over 25 km s$^{-1}$. While there are counterexamples to this pattern, such as the low-velocity groups of SCYA-104 and SCYA-72, mean velocities exceeding this speed only exist in $\sim$20\% of other associations, indicating these these high velocities are overrepresented in this set of tenuous populations. 

There are also three groups in this sample which reside more than 100 pc from the galactic plane, most notably SCYA-2 and SCYA-3, a pair of small associations with $200$ pc $<Z<300$ pc. Their locations high above the galactic plane are likely a side-effect of the high velocities seen in other groups, in which substantial vertical velocities near the galactic plane translate into positions high above the galactic plane at the peaks of their orbits' vertical oscillations. None of the groups with $|Z|>100$ pc have transverse velocities above 25 km s$^{-1}$, so they are likely currently near the peaks in their oscillations about the galactic midplane. As a result, the high velocities and high $Z$ values may be different manifestations of the same pattern affecting a majority of these populations, a pattern which may indicate a unique origin for at least a subset of these associations.

High group velocities likely imply high parent cloud velocities, which are common for material in the galactic halo \citep[e.g.,][]{Wakker97, Richter17}. These high-velocity clouds are thought to arise either through the direct infall of material from outside the galactic plane, or through the ejection of material from the galactic plane by a supernova which can then fall back to the galactic plane like a fountain \citep{Heitsch22}. While the contribution of these clouds to current star formation is not well-established \citep[e.g.,][]{Stark15}, recent surveys suggest that their presence extends close to the galactic disk, where interactions with dense gas in the galactic midplane may both compress the cloud and scoop up disk material, resulting in triggered star formation \citep[e.g.,][]{Lehner22,Fukui21}. If these groups do originate in these high-velocity clouds, their chemistry may provide a strong indication whether the clouds themselves originate in metal-rich disk material or metal-poor intergalactic gas, which has been an active topic of discussion in their study \citep[e.g.,][]{Marasco22}. These high-velocity associations could also be formed during or soon after ejection from a star-forming region, an origin that should be identifiable with trace back studies, which would be indicated by formation co-spatial with a large star-forming region. Nonetheless, a detailed census of the ages, orbits, chemistry, and populations of these associations will be necessary to assess their origins.

\section{Conclusion} \label{sec:concl}

Using the improved astrometry and photometry from Gaia DR3, we have identified more than $4\times10^4$ young stars, which trace numerous recent star forming events in the solar neighborhood. Many of the features detected have never been seen before and could provide insight into new patterns of star formation. Our key findings are as follows:

\begin{enumerate}
    \item We produced a new SPYGLASS Catalog of Young Associations (SCYA), which contains 116 young associations within 1 kpc. Of these associations, 10 are completely new discoveries, and a further 20 are known at least in part, but undergo significant redefinition of their extent. 
    \item Many populations in this new survey were found to have connecting structure linking them to other populations. This was most notably the case for the Orion Complex and Perseus OB2, which are connected by lower-density bridge populations including Taurus-Orion IV, which was identified in SPYGLASS-I. This could suggest a direct structural link early in their formation, which continues to manifest through these tenuous stellar populations
    \item We defined a substantial new star formation complex in Cep-Her, which has an age of approximately 30 Myr. Its spatial extent and population of thousands of stars makes it a potential analog to younger complexes like Sco-Cen at a more advanced stage of dynamical evolution. Along with a parallel structure, it also has an orientation directly perpendicular to the Radcliffe Wave, suggesting that the current star formation patterns which closely follow the spiral arms have only recently been active. 
    \item Many of the newly-discovered associations have unique dynamical properties, including a pair of associations (SCYA-2 and SCYA-3) located more than 200 pc from the galactic plane, and more associations with very high transverse velocities, including two which exceed 50 km s$^{-1}$. This could indicate a new and unique demographic of young associations with orbits inconsistent with most of the molecular gas in the galactic plane, potentially indicating an entirely different formation mechanism. 

\end{enumerate}

Our results show a wealth of unique structures that both hint to largely unknown processes, and suggest connections between structures that could reveal star formation patterns on a much larger scale. However, the large populations and extents of these structures necessitate collaboration to carefully assemble these small-scale associations into larger patterns. We hope that the extensive and accessible membership lists that we present here will help to facilitate these studies. 

\begin{acknowledgments}

RMPK is funded from the Heising-Simons Foundation. RMPK thanks the Texas Advanced Computing Center (TACC) at the University of Texas at Austin for access to their extensive computational resources, which were instrumental in facilitating our search for young populations in Gaia. RMPK also thanks Aaron Rizzuto, whose guidance and mentorship was essential to the development of the SPYGLASS program. The authors also thank Luke Bouma, whose helpful comments improved the clarity and content of this paper. 

\end{acknowledgments}

%

\vspace{5mm}
\facilities{Gaia}


\software{Astropy \citep{Astropy13},
Matplotlib \citep{Hunter07}, pandas \citep{Pandas20}, Numpy \citep{numpy20}
         }

\bibliography{spyglass4}{}
\bibliographystyle{aasjournal}



\appendix

\section{Older Populations} \label{app:old}

\input{oldpops.tex}

In Table \ref{tab:oldpopmems} we provide stellar populations associated with older populations, which were identified in this publication due to either a strong binary sequence or contamination from a molecular cloud in the foreground. The properties we include are the same as those we include in Table \ref{tab:allpops}, with the exception of our $P_{mem}$ measurements, which cannot be reliably provided without a high-confidence young population in the sample. We also provide a summary of basic properties for each population, similar to what we provide in Table \ref{tab:groupstats} for the young populations. 

\section{Circinus} \label{app:circinus}

\begin{figure}
  \centering 
  \includegraphics[width=7cm]{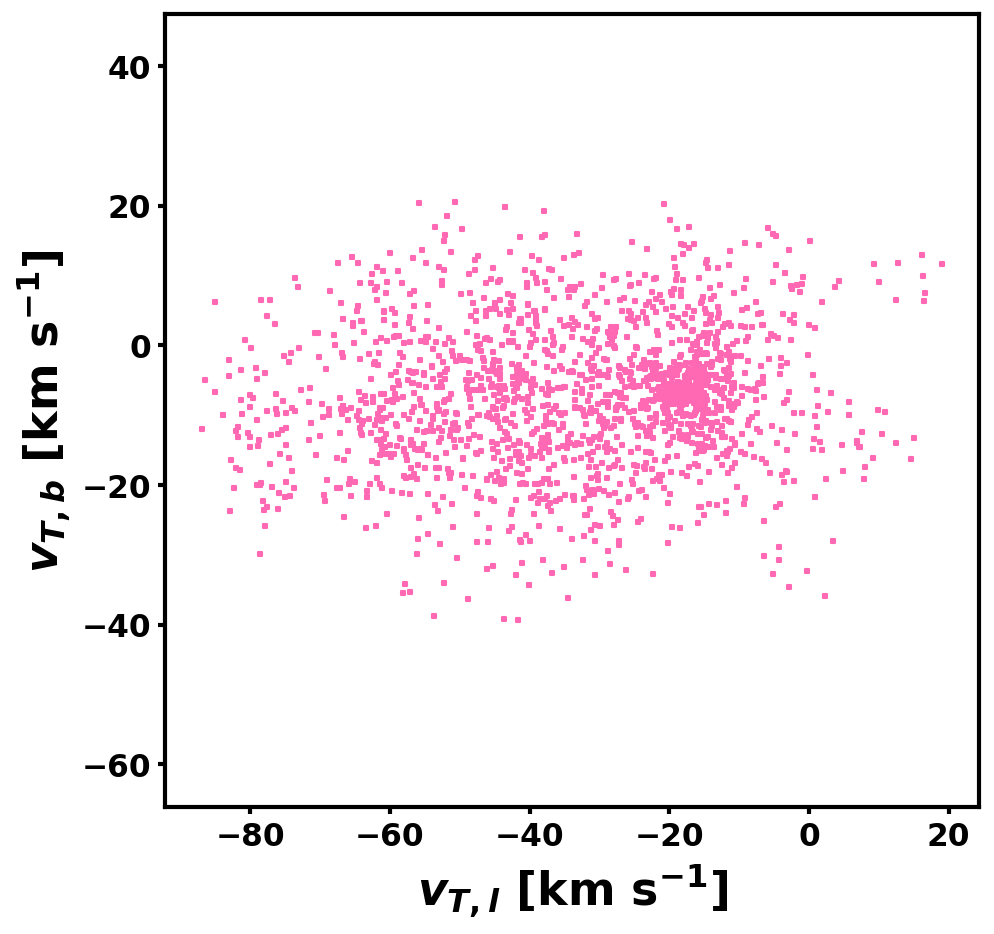}\hfill 
  \caption{The transverse velocity distribution for SCYA-6 (Circinus), which has high enough local reddening that a large sample of non-member field stars is included in the founding population, which follow a much broader velocity distribution compared to the group itself. Circinus is the only severe example of this, however it demonstrates that in the most distant and most heavily-reddened groups, harsher $P_{mem}$ restrictions which exclude some founding members may be desirable for future analysis. This velocity distribution can also be viewed in the online-only version of figure \ref{fig:vtassocsmap}.}
  \label{fig:vtcircinus}
\end{figure}

We plot the Circinus complex in transverse velocity space in Figure \ref{fig:vtcircinus}. This group is badly enough reddened that a substantial component of the local field was identified as young founding members, which assume the field velocity distribution. The field velocity distribution can be excluded from the Circinus sample by restricting further in $P_{mem}$, and allowing for the removal of founding members that fail these restrictions. The distribution of a group's members in transverse velocity space is typically a useful indicator of whether further restriction of the sample is necessary for future studies. This is, however, the only group with a significant component of its population following a field-like velocity distribution, demonstrating the velocity coherence of the other groups.

\end{document}

%% file: Memberlist.tex
\begin{deluxetable*}{ccccccccccccc}
\tablecolumns{13}
\tablewidth{0pt}
\tabletypesize{\scriptsize}
\tablecaption{Candidate members of young associations, sorted by their SCYA group ID. We include both members used to identify the population, and members of the extended population with less certain youth, provided that they have $P_{mem}>0.05$. We include the Gaia ID, numerous basic properties, and flags used to assess the Gaia observation quality and membership likelihood of each member.}
\label{tab:allpops}
\tablehead{
\colhead{SCYA} &
\colhead{Gaia DR3 ID} &
\colhead{RA} &
\colhead{Dec} &
\colhead{$d$} &
\colhead{$m_G$} &
\colhead{$G_{BP}-G_{RP}$} &
\colhead{$A$\tablenotemark{a}} &
\colhead{$P$\tablenotemark{b}} &
\colhead{$\pi/\sigma_{\pi}$} &
\colhead{$P_{\tau<50 Myr}$} &
\colhead{$P_{mem}$} &
\colhead{$F$\tablenotemark{c}} \\
\colhead{} &
\colhead{} &
\colhead{(deg)} &
\colhead{(deg)} &
\colhead{(pc)} &
\colhead{} &
\colhead{} &
\colhead{} &
\colhead{} &
\colhead{} &
\colhead{} &
\colhead{} &
\colhead{}
}
\startdata
    1 &  6086677607216395264 &  200.1376 & -47.2830 &   585.1 &  14.47 &  1.40 &                 1 &                 1 &     73.2 &   0.350 &  0.000 &         1 \\
    1 &  6088255921798929280 &  199.2986 & -44.2373 &   778.3 &  15.53 &  1.53 &                 1 &                 1 &     35.6 &   0.708 &  0.000 &         1 \\
    1 &  6097169628200000640 &  210.9157 & -44.2703 &   669.1 &  20.26 &  1.00 &                 1 &                 1 &      3.1 &      &  0.080 &         0 \\
    1 &  6097286313872329600 &  211.9991 & -43.8656 &   658.0 &  20.27 &  1.96 &                 1 &                 0 &      2.7 &      &  0.092 &         0 \\
    1 &  6097652966642631168 &  212.8071 & -43.3230 &   666.6 &  17.21 &  1.95 &                 1 &                 1 &     16.9 &   0.002 &  0.084 &         0 \\
    1 &  6097678221050944512 &  212.5070 & -43.3811 &   644.2 &  18.81 &  2.67 &                 1 &                 1 &      6.3 &   0.006 &  0.052 &         0 \\
    1 &  6097710347401574784 &  212.9618 & -43.0180 &   673.8 &  20.04 &  1.67 &                 1 &                 0 &      3.8 &      &  0.075 &         0 \\
    1 &  6097742817358919680 &  213.6807 & -43.1084 &   649.2 &  16.05 &  1.45 &                 1 &                 1 &     28.2 &   0.000 &  0.051 &         0 \\
    1 &  6097784220841457536 &  214.1132 & -42.5274 &   660.8 &  20.79 &  1.12 &                 1 &                 0 &      2.3 &      &  0.052 &         0 \\
    1 &  6107859900585069440 &  208.0558 & -44.4360 &   697.7 &  20.44 &  1.00 &                 1 &                 0 &      2.5 &      &  0.073 &         0 \\
    1 &  6107886293162139776 &  208.5639 & -44.6361 &   695.0 &  16.77 &  1.68 &                 1 &                 1 &     17.9 &   0.001 &  0.066 &         0 \\
    1 &  6107915597718417664 &  209.2874 & -44.1924 &   647.8 &  19.71 &  2.54 &                 1 &                 1 &      3.9 &      &  0.052 &         0 \\
\enddata
\tablenotetext{a}{The boolean solution to the astrometric quality cut, which is based on the unit weight error. 1 passes, 0 fails.}
\tablenotetext{b}{The boolean solution to the photometric quality cut, which is based on the BP/RP flux excess factor. 1 passes, 0 fails.}
\tablenotemark{c}{indicates whether the star was part of the robustly young stellar population used to identify the group in the clustering stage.}
\vspace*{0.1in}
\end{deluxetable*}

%% file: AllGroups.tex
\startlongtable
\begin{deluxetable*}{cccccccccccccccc}
\tablecolumns{16}
\tablewidth{0pt}
\tabletypesize{\scriptsize}
\tablecaption{Properties of the nearby clusters identified using HDBSCAN.} 
\label{tab:groupstats}
\tablehead{
\colhead{SCYA} &
\colhead{Name\tablenotemark{a}\tablenotemark{b}} &
\colhead{N\tablenotemark{c}} &
\colhead{RA} &
\colhead{Dec} &
\colhead{l} &
\colhead{b} &
\colhead{$D_{sky}$\tablenotemark{d}} &
\colhead{d} &
\colhead{$\mu_{RA}$} &
\colhead{$\mu_{Dec}$} &
\colhead{$V_{T, l}$} &
\colhead{$V_{T, b}$} &
\colhead{$\sigma_{V_T}$\tablenotemark{e}} &
\colhead{Age} \\
\colhead{} &
\colhead{} &
\colhead{} &
\multicolumn{2}{c}{(deg)} &
\multicolumn{2}{c}{(deg)} &
\colhead{(deg)} &
\colhead{(pc)} &
\multicolumn{2}{c}{(mas/yr)} &
\multicolumn{2}{c}{(km/s)} &
\colhead{(km/s)} &
\colhead{(Myr)}
}
\startdata
    1 &       NGC 5367 &    30 &  207.0 & -40.0 &  315.4 &  21.2 &      9.7 $\times$ 4.2 &  662 $\pm$ 53 &  -9.8 &   -0.7 & -30.9 &   5.6 &  23.9 $\times$ 8.5 &    9 \\
    2 &   H-R 581       &    19 &  173.7 &  16.9 &  238.7 &  69.6 &     10.9 $\times$ 3.7 &  309 $\pm$ 12 &  -9.1 &   -1.5 &  -5.4 & -12.7 &   2.3 $\times$ 0.9 &   21 \\
    3 &   H-R 3321      &    13 &  158.4 &  14.1 &  227.7 &  55.6 &      7.7 $\times$ 3.7 &  273 $\pm$ 16 &  -9.1 &   -1.8 &  -3.5 & -11.6 &   1.6 $\times$ 1.3 &   15 \\
    4 &       NGC 2169 &    22 &   92.1 &  14.0 &  195.5 &  -2.8 &      3.2 $\times$ 1.1 &  933 $\pm$ 37 &  -1.1 &   -1.7 &   4.6 &  -8.2 &   9.1 $\times$ 5.1 &   16 \\
    5 &        $cP595$ &    20 &   90.0 &  18.5 &  191.1 &  -2.2 &      5.2 $\times$ 2.4 &  593 $\pm$ 29 &  -0.9 &   -6.4 &  14.6 & -11.2 &   7.2 $\times$ 3.1 &   15 \\
    6 &       Circinus &  1756 &  226.5 & -60.9 &  318.3 &  -2.5 &      4.6 $\times$ 2.7 &  895 $\pm$ 65 &  -5.2 &   -4.8 & -30.0 &  -7.1 &  40.5 $\times$ 19.5 &    5 \\
    7 &            NAS &   160 &  250.5 & -46.6 &  338.7 &  -1.7 &      5.2 $\times$ 2.8 &  945 $\pm$ 42 &  -0.2 &   -3.1 & -11.2 &  -8.8 &  18.1 $\times$ 9.1 &    5 \\
    8 &                &    10 &  256.3 & -46.1 &  341.3 &  -2.9 &      2.4 $\times$ 1.1 &  960 $\pm$ 12 &  -3.0 &   -6.6 & -33.3 &  -7.6 &   6.1 $\times$ 5.4 &    7 \\
    9 &   $fT285/P513$ &    46 &   22.6 &  64.4 &  127.3 &   1.8 &      3.1 $\times$ 2.9 &  903 $\pm$ 40 &  -0.5 &   -1.5 &  -1.1 &  -6.6 &   6.9 $\times$ 4.4 &    6 \\
   10 &                &    14 &   91.1 & -64.5 &  274.1 & -29.4 &      8.2 $\times$ 7.5 &  406 $\pm$ 35 &  -6.8 &   14.5 & -27.9 & -12.9 &   3.7 $\times$ 2.4 &   27 \\
   11 &   H-R 3562     &    10 &  137.5 & -48.7 &  270.0 &  -0.5 &      1.0 $\times$ 0.1 &  688 $\pm$ 12 &  -7.6 &    3.6 & -25.8 & -10.3 &   2.1 $\times$ 2.0 &   11 \\
   12 &        IC 2395 &   123 &  130.5 & -48.0 &  266.5 &  -3.6 &      1.6 $\times$ 0.9 &  698 $\pm$ 23 &  -4.5 &    3.4 & -18.0 &  -4.8 &   2.5 $\times$ 1.5 &    8 \\
   13 &           P653 &    36 &  123.1 & -39.4 &  256.2 &  -3.4 &      2.7 $\times$ 1.5 &  858 $\pm$ 58 &  -4.1 &    3.9 & -22.5 &  -5.5 &   3.9 $\times$ 2.9 &    8 \\
   14 &           P676 &    30 &  138.8 & -45.6 &  268.7 &   1.9 &      1.8 $\times$ 1.1 &  738 $\pm$ 12 &  -8.9 &    4.1 & -32.0 & -12.2 &   3.8 $\times$ 3.1 &    2 \\
   15 &        $fP674$ &    87 &  137.3 & -47.6 &  269.2 &   0.0 &      2.5 $\times$ 1.5 &  838 $\pm$ 35 &  -7.7 &    5.3 & -35.7 &  -8.8 &  12.8 $\times$ 6.5 &    6 \\
   16 &        BH 56 &    26 &  134.6 & -43.1 &  264.5 &   1.7 &      0.9 $\times$ 0.6 &  895 $\pm$ 20 &  -5.6 &    5.3 & -32.4 &  -3.1 &   7.7 $\times$ 1.5 &   10 \\
   17 &         Vela-C &   252 &  131.1 & -41.2 &  261.4 &   1.0 &      2.5 $\times$ 0.7 &  914 $\pm$ 31 &  -5.6 &    3.9 & -28.1 &  -8.6 &   4.8 $\times$ 3.1 &    3 \\
   18 &         $fT51$ &    75 &   37.4 &  73.0 &  130.1 &  11.6 &      4.1 $\times$ 1.4 &  870 $\pm$ 66 &  -1.4 &   -1.5 &  -3.0 &  -7.9 &   6.4 $\times$ 6.2 &    5 \\
   19 &       Theia 83 &    65 &   89.7 & -14.4 &  220.3 & -17.8 &      3.3 $\times$ 2.4 &  656 $\pm$ 29 &  -1.9 &    1.5 &  -7.0 &  -3.8 &   3.5 $\times$ 2.2 &    6 \\
   20 &         VDB 80 &    28 &   96.9 & -10.0 &  219.3 &  -9.8 &      1.8 $\times$ 0.3 &  917 $\pm$ 34 &  -3.2 &    0.8 &  -9.6 & -10.9 &   4.1 $\times$ 3.1 &    3 \\
   21 &         Mon R2 &    75 &   92.1 &  -8.6 &  215.6 & -12.8 &      4.0 $\times$ 2.1 &  821 $\pm$ 27 &  -3.0 &    1.1 &  -9.1 &  -8.4 &   8.1 $\times$ 3.0 &    4 \\
   22 &           Cone &   287 &  100.0 &   9.8 &  202.8 &   1.9 &      2.6 $\times$ 1.4 &  688 $\pm$ 38 &  -2.0 &   -3.9 &   8.7 & -11.8 &   5.2 $\times$ 4.2 &    5 \\
   23 &      Theia 761 &    19 &   84.7 &  26.3 &  181.4 &  -2.6 &      3.0 $\times$ 0.7 &  740 $\pm$ 27 &  -1.5 &   -4.7 &  11.2 & -12.9 &   4.1 $\times$ 2.4 &    5 \\
   24 &      Cep OB2/3 &  1028 &  330.4 &  61.5 &  104.5 &   4.4 &     10.4 $\times$ 4.1 &  866 $\pm$ 59 &  -1.8 &   -2.9 & -12.9 &  -5.7 &  13.6 $\times$ 7.2 &    3 \\
   25 &    $fT107/177$ &    32 &  105.9 & -52.4 &  262.7 & -19.4 &      9.6 $\times$ 4.2 &  529 $\pm$ 39 &  -7.7 &    9.4 & -28.0 & -10.0 &   4.3 $\times$ 2.9 &   18 \\
   26 &                &    12 &  255.5 & -21.2 &    0.1 &  14.0 &     12.9 $\times$ 6.4 &  218 $\pm$ 26 & -16.0 &  -47.7 & -49.6 & -15.5 &   6.3 $\times$ 4.0 &   27 \\
   27 &         $cT73$ &    16 &  274.8 &  12.5 &   41.0 &  12.6 &      5.5 $\times$ 1.4 &  590 $\pm$ 13 &  -1.0 &   -6.9 & -18.4 &  -5.7 &   5.3 $\times$ 4.1 &   24 \\
   28 &                &    11 &  173.8 & -65.6 &  296.3 &  -4.8 &      8.3 $\times$ 6.7 &  214 $\pm$ 14 & -47.2 &    5.3 & -46.2 & -12.2 &   8.2 $\times$ 4.4 &   48 \\
   29 &         Cocoon &    58 &  328.4 &  47.2 &   94.3 &  -5.5 &      1.4 $\times$ 0.8 &  749 $\pm$ 28 &  -2.8 &   -2.9 & -14.4 &  -1.2 &   4.5 $\times$ 4.3 &    3 \\
   30 &         UPK 88 &    17 &  309.4 &  22.0 &   65.1 & -11.7 &      4.3 $\times$ 3.1 &  300 $\pm$ 21 &  10.0 &   -5.2 &   1.9 & -15.6 &   3.7 $\times$ 2.6 &   22 \\
   31 &          $fT6$ &    58 &   10.6 &  62.1 &  121.9 &  -0.7 &      5.2 $\times$ 2.5 &  527 $\pm$ 24 &   3.2 &   -2.2 &   7.6 &  -5.9 &   6.2 $\times$ 4.0 &    3 \\
   32 &    $fP533/548$ &    42 &   51.9 &  42.4 &  150.5 & -12.3 &     12.1 $\times$ 4.3 &  408 $\pm$ 28 &   5.5 &   -7.3 &  16.2 &  -5.9 &   7.8 $\times$ 2.5 &   20 \\
   33 &       Theia 19 &    54 &    3.4 &  66.1 &  119.1 &   3.4 &      2.8 $\times$ 1.8 &  702 $\pm$ 20 &   2.3 &   -2.4 &   6.3 &  -8.8 &   9.0 $\times$ 4.0 &    3 \\
   34 &      H-R 366 &    20 &  131.1 & -59.3 &  275.8 &  -9.9 &      4.2 $\times$ 1.4 &  523 $\pm$ 20 &  -7.8 &    6.9 & -25.0 &  -5.4 &   2.3 $\times$ 1.8 &   16 \\
   35 &                &    11 &  246.2 &  38.2 &   61.4 &  44.4 &     18.3 $\times$ 6.1 &  173 $\pm$ 11 &  -9.2 &   -3.9 &  -2.9 &   7.4 &   6.4 $\times$ 2.7 &   29 \\
   36 &           P505 &    14 &   12.3 &  50.8 &  122.5 & -12.0 &      2.0 $\times$ 1.1 &  355 $\pm$ 6  &   5.0 &   -1.3 &   8.4 &  -2.5 &   2.8 $\times$ 1.1 &    7 \\
   37 &         $cT58$ &    15 &  131.9 & -39.0 &  260.0 &   2.8 &      2.2 $\times$ 1.6 &  577 $\pm$ 29 &  -9.1 &    4.5 & -25.0 & -11.4 &   3.8 $\times$ 1.0 &   34 \\
   38 &       NGC 1579 &    41 &   67.5 &  35.3 &  165.3 &  -9.0 &      3.0 $\times$ 1.3 &  522 $\pm$ 25 &   2.3 &   -5.1 &  13.0 &  -4.4 &   3.5 $\times$ 2.9 &    1 \\
   39 &      Theia 105 &    17 &  138.7 & -45.5 &  268.5 &   2.9 &      3.3 $\times$ 2.2 &  505 $\pm$ 13 &  -9.3 &    0.0 & -15.4 & -15.5 &   3.8 $\times$ 1.4 &   15 \\
   40 &      Dolidze 8 &    37 &  305.0 &  41.0 &   78.5 &   2.7 &      1.4 $\times$ 0.6 &  969 $\pm$ 27 &  -2.5 &   -6.3 & -30.2 &  -6.8 &   4.1 $\times$ 2.6 &    2 \\
   41 &        $cP403$ &    19 &  307.5 &  40.1 &   78.9 &   0.7 &      2.3 $\times$ 0.7 &  940 $\pm$ 23 &  -1.9 &   -4.0 & -19.2 &  -3.5 &   3.6 $\times$ 2.3 &    2 \\
   42 &    Lacerta OB1 &   525 &  340.1 &  40.4 &   97.1 & -15.8 &     12.0 $\times$ 2.8 &  470 $\pm$ 43 &  -1.6 &   -4.6 &  -8.7 &  -6.9 &   7.8 $\times$ 3.8 &   10 \\
   43 &                &    16 &   31.7 &  66.4 &  130.4 &   4.7 &      5.9 $\times$ 2.9 &  291 $\pm$ 14 &  10.4 &   -8.2 &  16.8 &  -6.5 &   2.3 $\times$ 1.6 &   22 \\
   44 &        $cT624$ &    24 &  257.1 & -37.7 &  348.1 &   1.7 &      6.2 $\times$ 4.3 &  407 $\pm$ 15 &  -2.7 &   -7.6 & -14.8 &  -4.5 &   2.8 $\times$ 2.0 &   41 \\
   45 &         Cr 132 &    38 &  107.8 & -30.0 &  242.0 &  -9.2 &      5.3 $\times$ 1.9 &  612 $\pm$ 21 &  -4.3 &    3.9 & -16.0 &  -6.5 &   2.9 $\times$ 1.6 &   22 \\
   46 &        VDB 83 &    10 &  100.0 & -27.2 &  236.5 & -14.3 &      1.8 $\times$ 0.2 &  898 $\pm$ 9  &  -2.8 &    3.3 & -17.6 &  -5.5 &   1.2 $\times$ 1.0 &   13 \\
   47 &          CaMaS &   146 &  105.2 & -24.9 &  235.9 &  -9.3 &      3.7 $\times$ 2.1 &  809 $\pm$ 54 &  -3.1 &    3.5 & -17.3 &  -5.0 &   2.2 $\times$ 1.9 &   16 \\
   48 &             M6 &    17 &  265.0 & -32.3 &  356.5 &  -0.7 &      2.1 $\times$ 0.6 &  464 $\pm$ 14 &  -1.4 &   -5.8 & -12.3 &  -4.1 &   3.0 $\times$ 1.3 &   42 \\
   49 &    H-R 4379 &    15 &  120.5 & -44.7 &  259.9 &  -7.5 &      2.3 $\times$ 1.2 &  569 $\pm$ 21 &  -6.3 &    7.3 & -25.7 &  -4.4 &   1.6 $\times$ 1.4 &   24 \\
   50 &           P441 &    16 &  315.3 &  56.6 &   95.2 &   6.1 &      3.3 $\times$ 1.8 &  484 $\pm$ 7  &   0.2 &   -1.4 &  -1.9 &  -2.4 &   3.1 $\times$ 2.3 &    3 \\
   51 &        $cP418$ &    14 &  314.4 &  44.2 &   85.2 &  -0.9 &      1.8 $\times$ 0.7 &  881 $\pm$ 14 &  -2.3 &   -3.4 & -17.3 &  -2.7 &   4.7 $\times$ 3.3 &    1 \\
   52 &        Theia 1 &    95 &  315.6 &  50.8 &   91.0 &   2.7 &      2.7 $\times$ 1.4 &  587 $\pm$ 37 &   1.2 &   -3.5 &  -5.1 &  -9.1 &   4.2 $\times$ 2.6 &    2 \\
   53 &             NA &   281 &  314.3 &  44.5 &   85.3 &  -0.7 &      4.3 $\times$ 2.0 &  755 $\pm$ 46 &  -0.7 &   -3.0 &  -9.4 &  -5.5 &   7.2 $\times$ 4.3 &    3 \\
   54 &      Theia 232 &    21 &   16.2 &  51.2 &  125.1 & -11.5 &      4.6 $\times$ 2.9 &  323 $\pm$ 17 &  15.4 &   -6.7 &  23.9 &  -8.9 &   2.2 $\times$ 1.3 &   30 \\
   55 &       NGC 3228 &    19 &  155.1 & -51.8 &  280.7 &   4.3 &      2.2 $\times$ 0.8 &  474 $\pm$ 10 & -14.9 &   -0.6 & -27.4 & -19.4 &   2.8 $\times$ 1.0 &   21 \\
   56 &             FH &    58 &   54.3 & -34.8 &  235.7 & -53.1 &    21.5 $\times$ 15.6 &  104 $\pm$ 15 &  34.5 &   -4.9 &   1.5 &  17.1 &   7.7 $\times$ 5.3 &   43 \\
   57 &            P25 &    28 &  104.6 & -66.5 &  277.2 & -24.2 &      3.8 $\times$ 3.4 &  410 $\pm$ 15 &  -2.1 &    7.5 & -15.3 &  -0.8 &   2.1 $\times$ 0.9 &   14 \\
   58 &                &    18 &   36.6 &  58.7 &  135.2 &  -1.6 &      4.7 $\times$ 3.5 &  290 $\pm$ 15 &  16.7 &  -11.3 &  26.8 &  -6.4 &   2.5 $\times$ 2.2 &   25 \\
   59 &           P369 &    11 &  290.1 &  11.3 &   46.3 &  -1.2 &      2.5 $\times$ 0.3 &  403 $\pm$ 7  &   2.5 &  -10.4 & -15.3 & -13.8 &   2.5 $\times$ 1.2 &    1 \\
   60 &      Theia 124 &    37 &  310.3 &  64.3 &   99.5 &  13.6 &      2.5 $\times$ 1.9 &  447 $\pm$ 16 &   4.4 &   -0.1 &   5.2 &  -7.7 &   2.2 $\times$ 1.5 &    3 \\
   61 &       ASCC 107 &    38 &  295.7 &  21.4 &   57.6 &  -1.3 &      2.2 $\times$ 1.2 &  862 $\pm$ 43 &  -0.3 &   -5.3 & -19.5 &  -9.9 &   4.6 $\times$ 2.8 &   10 \\
   62 &           P652 &   104 &  123.9 & -38.7 &  255.9 &  -2.3 &      2.1 $\times$ 2.0 &  587 $\pm$ 24 &  -4.9 &    5.2 & -19.9 &  -3.3 &   3.4 $\times$ 1.6 &    6 \\
   63 &   Cr 359 &    30 &  270.7 &   3.0 &   30.1 &  12.3 &      3.4 $\times$ 1.9 &  547 $\pm$ 16 &   0.7 &   -8.9 & -19.6 & -12.0 &   2.5 $\times$ 1.2 &   14 \\
   64 &      Tau-Ori 1 &    50 &   66.1 &  14.7 &  181.3 & -23.6 &      5.1 $\times$ 3.1 &  303 $\pm$ 17 &   4.6 &   -5.6 &  10.7 &   0.3 &   3.8 $\times$ 1.3 &   10 \\
   65 &      CMa North &    41 &  102.1 & -15.1 &  226.0 &  -7.7 &      5.9 $\times$ 4.2 &  178 $\pm$ 21 &  -0.0 &   -5.3 &   4.0 &  -1.7 &   3.3 $\times$ 1.9 &   30 \\
   66 &       Theia 72 &    38 &  101.7 &  -9.6 &  220.9 &  -5.3 &      5.7 $\times$ 1.5 &  273 $\pm$ 18 &  -1.0 &   -4.4 &   4.7 &  -3.8 &   2.0 $\times$ 1.4 &   30 \\
   67 &         UPK 73 &    68 &  293.4 &  22.0 &   57.3 &   1.3 &      3.1 $\times$ 1.0 &  647 $\pm$ 26 &   0.7 &   -6.4 & -15.8 & -10.9 &   8.9 $\times$ 4.9 &    4 \\
   68 &        H-R 5089 &    13 &  311.7 &  48.5 &   87.5 &   3.3 &      7.5 $\times$ 6.0 &  158 $\pm$ 9  &  12.0 &    2.9 &   7.6 &  -5.0 &   3.5 $\times$ 1.9 &   59 \\
   69 &            CFN &   197 &  330.0 &  75.2 &  113.3 &  17.4 &     15.8 $\times$ 5.5 &  174 $\pm$ 26 &  16.0 &    6.4 &  14.2 &  -2.9 &   4.5 $\times$ 1.8 &   21 \\
   70 &        $cT431$ &    22 &  277.8 &  -7.7 &   22.8 &   0.6 &      8.1 $\times$ 6.7 &  204 $\pm$ 16 &   0.5 &  -18.7 & -15.9 &  -8.9 &   3.0 $\times$ 2.5 &   11 \\
   71 &      Theia 148 &    17 &  285.3 &  22.1 &   53.8 &   7.8 &      2.3 $\times$ 1.4 &  577 $\pm$ 18 &   2.1 &   -5.2 & -10.6 & -11.4 &   1.6 $\times$ 1.1 &   17 \\
   72 &                &    12 &   65.8 &  -2.2 &  196.1 & -32.5 &     16.5 $\times$ 2.6 &  176 $\pm$ 9  &   3.0 &    6.0 &  -2.7 &   4.8 &   5.2 $\times$ 1.0 &   24 \\
   73 &        IC 4665 &    63 &  266.6 &   5.3 &   30.3 &  16.9 &      2.9 $\times$ 2.6 &  340 $\pm$ 28 &  -0.9 &   -8.6 & -13.1 &  -5.1 &   2.1 $\times$ 1.9 &   26 \\
   74 &           P344 &    11 &  282.1 &   0.9 &   33.8 &   1.5 &      2.7 $\times$ 1.2 &  563 $\pm$ 10 &   1.6 &   -6.4 & -13.4 & -11.3 &   3.0 $\times$ 1.6 &    4 \\
   75 &         Oph SE &    50 &  257.6 & -18.2 &    4.3 &  12.9 &      7.0 $\times$ 3.7 &  214 $\pm$ 19 &  -5.4 &  -11.5 & -12.6 &  -1.9 &   4.7 $\times$ 1.9 &   18 \\
   76 &         UPK 38 &    15 &  280.7 &  -1.6 &   30.5 &   1.1 &      2.6 $\times$ 0.7 &  558 $\pm$ 10 &  -1.2 &   -5.3 & -14.1 &  -3.8 &   1.9 $\times$ 1.6 &    5 \\
   77 &       NGC2451A &    66 &  116.1 & -39.0 &  253.2 &  -7.7 &     11.9 $\times$ 3.0 &  189 $\pm$ 14 & -21.4 &   15.8 & -21.9 & -10.1 &   2.9 $\times$ 1.7 &   42 \\
   78 &    Aquila East &    25 &  298.9 &  -8.1 &   32.7 & -17.4 &     13.0 $\times$ 4.9 &  134 $\pm$ 6  &   8.6 &  -26.3 & -12.7 & -12.4 &   3.9 $\times$ 1.9 &   18 \\
   79 &       Theia 98 &    18 &  311.2 &  45.2 &   84.4 &   1.9 &      6.4 $\times$ 1.9 &  436 $\pm$ 15 &   3.6 &    0.1 &   4.3 &  -5.7 &   3.1 $\times$ 1.6 &   29 \\
   80 &           Iris &    94 &  319.8 &  68.9 &  106.0 &  13.6 &      9.9 $\times$ 2.8 &  336 $\pm$ 12 &   8.2 &   -1.3 &   8.1 & -11.0 &   6.1 $\times$ 1.7 &    8 \\
   81 &           P468 &    16 &  328.9 &  65.5 &  106.0 &   8.6 &      2.7 $\times$ 0.9 &  448 $\pm$ 11 &   5.7 &   -0.4 &   8.9 &  -8.2 &   1.2 $\times$ 1.1 &   12 \\
   82 &        Serpens &   687 &  277.5 &  -3.5 &   27.2 &   3.2 &      6.9 $\times$ 3.6 &  433 $\pm$ 47 &   2.0 &   -8.5 & -13.1 & -11.0 &   8.4 $\times$ 6.1 &    5 \\
   83 &      Cep Flare &   606 &  339.4 &  66.5 &  111.7 &   5.7 &     19.3 $\times$ 8.7 &  365 $\pm$ 29 &   6.6 &   -1.5 &   9.7 &  -7.3 &   7.0 $\times$ 4.3 &   12 \\
   84 &      Alessi 20 &   693 &  356.9 &  58.3 &  114.2 &  -3.7 &      8.6 $\times$ 3.7 &  419 $\pm$ 26 &   7.6 &   -2.4 &  12.8 &  -8.1 &   5.4 $\times$ 2.7 &   11 \\
   85 &                &    10 &  232.2 & -65.6 &  318.3 &  -7.6 &      4.5 $\times$ 3.6 &  232 $\pm$ 7  &  -9.1 &  -23.7 & -22.8 & -15.3 &   2.9 $\times$ 1.4 &   22 \\
   86 &       ASCC 123 &    15 &  340.3 &  53.9 &  104.2 &  -3.9 &      4.9 $\times$ 1.2 &  226 $\pm$ 9  &  12.5 &   -1.7 &  11.0 &  -7.9 &   2.0 $\times$ 1.2 &   44 \\
   87 &      Theia 136 &    16 &  329.6 &  52.1 &   97.9 &  -2.8 &      4.1 $\times$ 3.3 &  267 $\pm$ 8  &  10.3 &   -1.0 &   9.6 &  -8.7 &   4.3 $\times$ 1.3 &   80 \\
   88 &        $fT133$ &    25 &  356.4 &  65.0 &  116.1 &   3.2 &      9.8 $\times$ 6.7 &  184 $\pm$ 14 &  23.4 &    2.0 &  20.7 &  -2.5 &   3.3 $\times$ 2.3 &   43 \\
   89 &       Theia 38 &   447 &  291.1 &  22.1 &   56.1 &   3.2 &      3.8 $\times$ 1.8 &  493 $\pm$ 24 &   1.2 &   -6.0 & -11.3 &  -9.1 &   4.2 $\times$ 2.1 &    7 \\
   90 &      Roslund 5 &    15 &  302.4 &  32.8 &   70.5 &  -0.2 &      3.2 $\times$ 2.5 &  375 $\pm$ 9  &   3.9 &   -0.9 &   2.5 &  -6.8 &   2.1 $\times$ 1.1 &   80 \\
   91 &        Cep OB6 &    19 &  333.0 &  56.0 &  101.8 &  -0.3 &      8.0 $\times$ 1.3 &  241 $\pm$ 8  &  16.7 &    4.7 &  18.9 &  -5.9 &   2.8 $\times$ 0.7 &   59 \\
   92 &     Haffner 13 &    26 &  115.4 & -30.2 &  245.2 &  -3.6 &      1.7 $\times$ 1.0 &  551 $\pm$ 20 &  -6.2 &    6.0 & -21.6 &  -6.3 &   2.0 $\times$ 1.0 &   26 \\
   93 &        H-R 5495 &    35 &  120.2 & -31.5 &  248.4 &  -0.5 &      4.5 $\times$ 1.5 &  478 $\pm$ 13 &  -8.9 &    5.7 & -21.9 & -10.2 &   2.9 $\times$ 1.9 &   29 \\
   94 &        H-R 574 &    13 &  113.4 & -25.3 &  240.0 &  -2.4 &      2.4 $\times$ 1.6 &  445 $\pm$ 10 &  -7.8 &    5.8 & -18.8 &  -8.5 &   1.8 $\times$ 1.4 &   28 \\
   95 &     $fT96/236$ &    48 &  291.5 &  21.4 &   56.0 &   2.0 &      9.4 $\times$ 3.2 &  291 $\pm$ 29 &   1.3 &   -8.6 &  -9.4 &  -7.3 &   6.0 $\times$ 2.0 &   24 \\
   96 &        Cep-Her &  1164 &  285.7 &  36.7 &   66.9 &  12.9 &     33.6 $\times$ 7.1 &  337 $\pm$ 28 &   1.7 &   -2.7 &  -3.2 &  -3.5 &  14.9 $\times$ 3.1 &   29 \\
   97 &    H-R 2562    &    10 &    5.4 &  30.9 &  115.0 & -31.5 &      9.9 $\times$ 4.8 &  167 $\pm$ 5  &  14.5 &   -3.0 &  10.9 &  -4.1 &   1.3 $\times$ 0.6 &   33 \\
   98 &    H-R 568 &    11 &  135.2 & -55.3 &  273.6 &  -6.1 &      2.3 $\times$ 0.9 &  423 $\pm$ 11 & -10.7 &    7.6 & -26.0 &  -6.1 &   2.4 $\times$ 0.9 &   31 \\
   99 &      Theia 246 &    14 &  189.1 & -66.2 &  301.4 &  -3.4 &      2.3 $\times$ 1.3 &  412 $\pm$ 10 & -13.6 &   -3.1 & -26.1 &  -7.1 &   2.9 $\times$ 1.6 &   26 \\
  100 &                &    11 &  123.5 & -36.4 &  254.8 &  -1.3 &      5.7 $\times$ 2.0 &  230 $\pm$ 10 &  -5.4 &    3.8 &  -7.2 &  -2.4 &   2.6 $\times$ 0.7 &   29 \\
  101 &       H-R 593 &    12 &   86.1 &  -1.1 &  206.2 & -15.5 &      2.1 $\times$ 1.0 &  439 $\pm$ 8  &   1.6 &   -4.0 &   8.7 &  -0.9 &   2.8 $\times$ 1.0 &    1 \\
  102 &            P51 &    10 &  156.5 & -66.9 &  289.4 &  -8.4 &      3.1 $\times$ 1.8 &  365 $\pm$ 11 & -16.6 &    2.9 & -27.5 & -11.2 &   2.2 $\times$ 1.0 &   31 \\
  103 &        Sco-Cen &  8147 &  233.6 & -39.5 &  335.4 &  11.7 &    50.0 $\times$ 20.3 &  139 $\pm$ 24 & -17.1 &  -21.7 & -18.2 &  -6.3 &   5.2 $\times$ 3.8 &   17 \\
  104 &  H-R 2902   &    23 &   44.0 &  17.0 &  160.0 & -36.2 &     16.1 $\times$ 4.1 &  139 $\pm$ 8  &  11.7 &   -7.1 &   9.0 &   0.6 &   2.6 $\times$ 1.3 &   25 \\
  105 &        $cT120$ &    10 &  225.1 & -66.2 &  315.6 &  -6.9 &      4.1 $\times$ 3.4 &  272 $\pm$ 10 & -13.0 &  -13.7 & -23.4 &  -7.7 &   1.5 $\times$ 1.3 &   42 \\
  106 &     $fT94/P82$ &    43 &  226.4 & -80.5 &  308.6 & -19.7 &     13.2 $\times$ 4.6 &  188 $\pm$ 29 &  -5.4 &  -14.1 & -11.0 &  -9.9 &   2.1 $\times$ 1.2 &   29 \\
  107 &        Per OB3 &   281 &   70.0 &  29.9 &  170.3 &  -5.5 &     37.2 $\times$ 8.9 &  180 $\pm$ 23 &  16.4 &  -23.3 &  24.8 &  -3.3 &   8.3 $\times$ 3.2 &   58 \\
  108 &         Taurus &   640 &   74.0 &  23.0 &  179.7 & -15.6 &    22.8 $\times$ 13.2 &  143 $\pm$ 25 &   6.9 &  -20.6 &  15.1 &  -5.5 &   9.7 $\times$ 6.2 &   11 \\
  109 &   Carina-Musca &   253 &  172.8 & -70.1 &  296.4 & -10.1 &     16.6 $\times$ 5.0 &  268 $\pm$ 43 & -22.9 &   -1.6 & -26.9 & -11.9 &   5.1 $\times$ 2.2 &   28 \\
  110 &         $cT75$ &    13 &  194.3 & -71.7 &  303.4 &  -8.9 &      4.0 $\times$ 1.7 &  382 $\pm$ 13 & -14.0 &   -6.9 & -25.1 & -11.5 &   1.6 $\times$ 1.0 &   17 \\
  111 &    $fT75/P144$ &    97 &  223.8 & -66.4 &  314.9 &  -6.6 &     13.3 $\times$ 2.2 &  395 $\pm$ 41 &  -7.2 &  -11.1 & -22.1 & -12.1 &   3.8 $\times$ 1.3 &   27 \\
  112 &       Theia 78 &    23 &   80.4 &  -4.4 &  206.8 & -22.5 &      3.5 $\times$ 2.1 &  318 $\pm$ 9  &   0.1 &    5.2 &  -6.9 &   3.8 &   1.9 $\times$ 1.4 &   16 \\
  113 &          Tr 10 &   420 &  144.2 & -52.6 &  276.0 &  -0.2 &    19.4 $\times$  2.7 &  410 $\pm$ 28 & -14.6 &    4.9 & -27.3 & -11.7 &   3.1 $\times$ 2.3 &   47 \\
  114 &           Vela &  5413 &  120.4 & -45.6 &  260.4 &  -8.3 &     22.0 $\times$ 8.3 &  359 $\pm$ 62 &  -6.7 &    8.3 & -18.3 &  -2.0 &   9.3 $\times$ 5.2 &   19 \\
  115 &        Ori-Per &  9147 &   83.2 &  -2.3 &  205.7 & -18.5 &    30.9 $\times$ 10.4 &  354 $\pm$ 50 &   1.1 &   -0.8 &   1.7 &   0.5 &   8.1 $\times$ 6.2 &    9 \\
  116 &  $\lambda$ Ori &   703 &   83.6 &   9.8 &  195.1 & -12.0 &      3.4 $\times$ 2.9 &  391 $\pm$ 13 &   1.4 &   -2.2 &   4.7 &   0.2 &   3.3 $\times$ 1.9 &    6 \\
\enddata
\tablenotetext{a}{We use shorthand markers to indicate connections to known groups in cases where known populations provide either only fragmentary coverage of our groups or have a very different extent. We indicate the relationship of the known group to our group as either fragmentary component of our group ($f$) or a larger structure that contains our group, ($c$). This is followed by the Group, either $T$ for Theia or $P$ for \citet{Prisinzano22}. For example, a group containing fragmentary components of Theia 23 would be recorded as $fT23$. Groups identified by an ``H-R'' ID are from the \citet{Hunt23} catalog.}
\tablenotetext{b}{Further explanations for components of complexes and shorthand names that lack specificity -- Circinus: associated with Circinus Moecular cloud, ASCC 79; NAS: NGC 6250 and NGC 6178; Vela-C: Cr 197, Gum 17; Mon R2: Monoceros R2 Cloud including NGC 2183; Cone: Cone Nebula Complex and NGC 2264; Cocoon: IC 5146 and Cocoon Nebula Complex; CaMaS: part of P633, sections of Theia 86 and 87; NA: North America Nebula Complex; Iris: NGC 7023 and Iris Nebula Complex; Cep Flare: Includes traditional Cepheus Flare region \citep[e.g.,][]{Szilagyi21} as well as ASCC 127 and most of the southern half of Theia 57; Cep-Her: $\delta$ Lyr, RSG-5, SPYGLASS-I groups Lyra, Cerberus, Cepheus-Cygnus; Tr 10: full Trumpler 10 complex, similar in extent to Theia 143; Orion-Perseus: Orion Nebula Complex, Perseus OB2, SPYGLASS-I groups Monocerous Southwest, Taurus-Orion II and IV. }
\tablenotetext{c}{Number of stars in the founding young stellar population.}
\tablenotetext{d}{On-sky spatial extent in galactic l/b, in the form of the RMS in major axis $\times$ minor axis, when fit with a bivariate Gaussian.}
\tablenotetext{e}{The extent of the velocity distribution, in the form of the RMS in semi-major axis $\times$ semi-minor axis, when fit with a bivariate Gaussian.}
\vspace*{0.1in}
\end{deluxetable*}

%% file: oldpops.tex
\begin{deluxetable*}{cccccccccccc}
\tablecolumns{12}
\tablewidth{0pt}
\tabletypesize{\scriptsize}
\tablecaption{Candidate members of older associations and their parent groups. We include both members used to identify the population, and members of the extended population. We include the Gaia ID, numerous basic properties, and flags used to assess the Gaia observation quality.}
\label{tab:oldpopmems}
\tablehead{
\colhead{Gaia DR3 ID} &
\colhead{Group \tablenotemark{a}} &
\colhead{RA} &
\colhead{Dec} &
\colhead{$d$} &
\colhead{$m_G$} &
\colhead{$G_{BP}-G_{RP}$} &
\colhead{$A$\tablenotemark{b}} &
\colhead{$P$\tablenotemark{c}} &
\colhead{$\pi/\sigma_{\pi}$} &
\colhead{$P_{\tau<50 Myr}$} &
\colhead{$F$\tablenotemark{d}} \\
\colhead{} &
\colhead{} &
\colhead{(deg)} &
\colhead{(deg)} &
\colhead{(pc)} &
\colhead{} &
\colhead{} &
\colhead{} &
\colhead{} &
\colhead{} &
\colhead{} &
\colhead{}
}
\startdata
 3073102052940845056 &      1 &  127.7206 &  -1.6946 &  192.9 &  14.31 &  1.81 &                 1 &                 1 &    247.1 &   0.001 &         0 \\
 3074406138155856384 &      1 &  133.5122 &  -0.1955 &  183.1 &  17.27 &  2.79 &                 1 &                 1 &     55.3 &   0.000 &         0 \\
 3074679126277191808 &      1 &  132.6266 &   0.2659 &  201.2 &  16.32 &  2.64 &                 1 &                 1 &     90.7 &   0.010 &         0 \\
 3074695795045520896 &      1 &  133.6876 &   0.1658 &  192.9 &  18.13 &  2.99 &                 1 &                 1 &     36.0 &   0.000 &         0 \\
 3075091481792346624 &      1 &  131.7143 &   0.3108 &  161.1 &  10.65 &  0.73 &                 1 &                 1 &    390.6 &   0.005 &         0 \\
 3075097941423221120 &      1 &  131.2496 &   0.2269 &  195.9 &  15.94 &  2.60 &                 0 &                 0 &     17.5 &     NaN &         0 \\
 3075341204075613824 &      1 &  130.6665 &   0.6511 &  157.5 &  16.57 &  2.75 &                 1 &                 1 &    103.0 &   0.002 &         0 \\
 3075341204076959488 &      1 &  130.6667 &   0.6501 &  165.1 &  18.28 &  3.52 &                 1 &                 1 &     34.3 &   0.002 &         0 \\
 3075696110108150528 &      1 &  131.2381 &   0.8175 &  169.1 &  10.74 &  0.74 &                 1 &                 1 &    360.9 &   0.007 &         0 \\
 3075905631498094080 &      1 &  132.0067 &   1.9072 &  157.4 &  16.74 &  2.60 &                 1 &                 1 &     84.0 &   0.001 &         0 \\
 3076421753423009152 &      1 &  126.7737 &  -0.7400 &  187.8 &  15.71 &  2.31 &                 1 &                 1 &    110.9 &   0.003 &         0 \\
 3076720919369814016 &      1 &  128.7206 &   0.0094 &  161.5 &  14.41 &  1.95 &                 1 &                 1 &    256.4 &   0.001 &         0 \\
\enddata
\tablenotetext{a}{The old group index. 1 is Praesepe, 2 is NGC 1342, 3 is NGC 6124, 4 is NGC 6997, 5 is MWSC 0243, 6 is Stock 2, and 7 is the Pleiades.}
\tablenotetext{b}{The boolean solution to the astrometric quality cut, which is based on the unit weight error. 1 passes, 0 fails.}
\tablenotetext{c}{The boolean solution to the photometric quality cut, which is based on the BP/RP flux excess factor. 1 passes, 0 fails.}
\tablenotemark{d}{indicates whether the star was part of the stellar population used to identify the group in the clustering stage (likely from the binary sequence).}
\vspace*{0.1in}
\end{deluxetable*}